\documentclass[12pt]{article}

\usepackage{amsmath}
\usepackage{graphicx}
\usepackage{xr}
\usepackage[font={sf}]{caption}
\usepackage{multirow}
\usepackage{tabularx}  % for 'tabularx' environment and 'X' column type
\author{Georg Meisl$^1$, Xiaoting Yang$^2$, Christopher M. Dobson$^1$, Sara Linse$^2$\\and Tuomas P. J. Knowles$^1$}
\date{}
\title{\textbf{A general reaction network unifies the aggregation behaviour of the A$\beta$42 peptide and its variants.}}
\begin{document}

\maketitle

\noindent $^1$Department of Chemistry, University of Cambridge, Lensfield Road, Cambridge CB2 1EW, UK $^2$Chemistry Department and Molecular Protein Science, Lund University, P. O. Box 124, SE221 00 Lund, Sweden\\

\begin{abstract}
\noindent The amyloid $\beta$ peptide (A$\beta$42), whose aggregation is associated with Alzheimer's disease, is an amphiphatic peptide with a high propensity to self-assemble.  A$\beta$42 has a net negative charge at physiological pH and modulations of intermolecular electrostatic interactions can significantly alter its aggregation behaviour. Variations in sequence and solution conditions lead to varied macroscopic behaviour, often resulting in a number of different mechanistic explanations for the aggregation of these closely related systems. Here we alter the electrostatic interactions governing the fibril aggregation kinetics by varying the ionic strength over an order of magnitude, which allows us to sample the space of different reaction mechanisms, and develop a minimal reaction network that explains the experimental kinetics under all the different conditions. We find that an increase in the ionic strength leads to an increased rate of surface catalysed nucleation over fragmentation and eventually to a saturation of this nucleation process.
More generally, this reaction network connects previously separate systems, such as mutants of A$\beta$42 and the wild type, on a continuous mechanistic landscape, thereby providing a unified picture of the aggregation mechanism of A$\beta$42 and the means of directly comparing the effects of intrinsic modifications of the peptide to those of simple electrostatic shielding. 
\end{abstract}

\section{Introduction}
Most functional proteins have a net charge under normal physiological conditions, which helps confer solubility\cite{ Lawrence2007,Kurnik2012,marti2000}, and is governed by the protein sequence and structure, as well as the solution conditions such as pH, salt concentration and the concentration of other charged species\cite{lindman2006a,silva2005,Kesvatera1996,kesvatera1999}.
The interactions involving charged and polar groups modulate properties such as solubility, stability and reaction rates \cite{Linse1988,Zhou2005,Matousek2007,Lindman2006b,Xue2009,Kurnik2012}. In addition to their importance in the functional interactions of proteins, electrostatic interactions play a key role in the formation of aberrant protein aggregates\cite{Vendruscolo2007,Chiti2006, Knowles2014, Schreiber2009}. In particular, charged proteins with embedded hydrophobic segments can be highly aggregation-prone and their assembly into amyloid fibrils is associated with Alzheimer's disease (the A$\beta$ peptide), Parkinson's disease (the protein $\alpha$-synuclein) and a range of other debilitating human diseases. The aggregation kinetics of these proteins are strongly influenced by electrostatic interactions and therefore depend on solution conditions and the presence of species able to shield charges\cite{Buell2013,Abelein2015}. 

Recent years have seen a significant advance in the mechanistic understanding of the aggregation of disease-associated proteins under controlled conditions in vitro \cite{Knowles2009, Cohen2011a, Meisl2016}. 
The mechanistic effects of variations in solution conditions, however, have often not been characterised in detail and therefore only the part of the overall reaction network relevant under a given set of conditions has been investigated. The individual systems under different conditions are not linked together into a continuous mechanistic picture. A more complete reaction network will be particularly important in vivo where aggregation-prone proteins are found in the presence of a large number of other molecules, which modulate their interactions.

Here, we present a method of sampling a large region of the reaction network of an aggregating system by modulating electrostatic interactions. This approach provides a means of altering the relative importance of different processes contributing to the overall reaction network and thereby allows the sampling of a broad range of macroscopic behaviour that can be explained by a single reaction network.
In the present work we investigate the aggregation kinetics of the 42-residue amyloid $\beta$ peptide, A$\beta$42, at different peptide and salt concentrations under quiescent conditions. We develop a model that quantitatively accounts for the observed lag times, kinetic profiles and peptide concentration dependences over the range of ionic strengths studied and rationalizes the interplay of the individual microscopic rates and their dependence on the magnitude of the electrostatic screening.

\section*{Results and Discussion}
\begin{figure}[h!]
	\centering
		\includegraphics[width=\columnwidth]{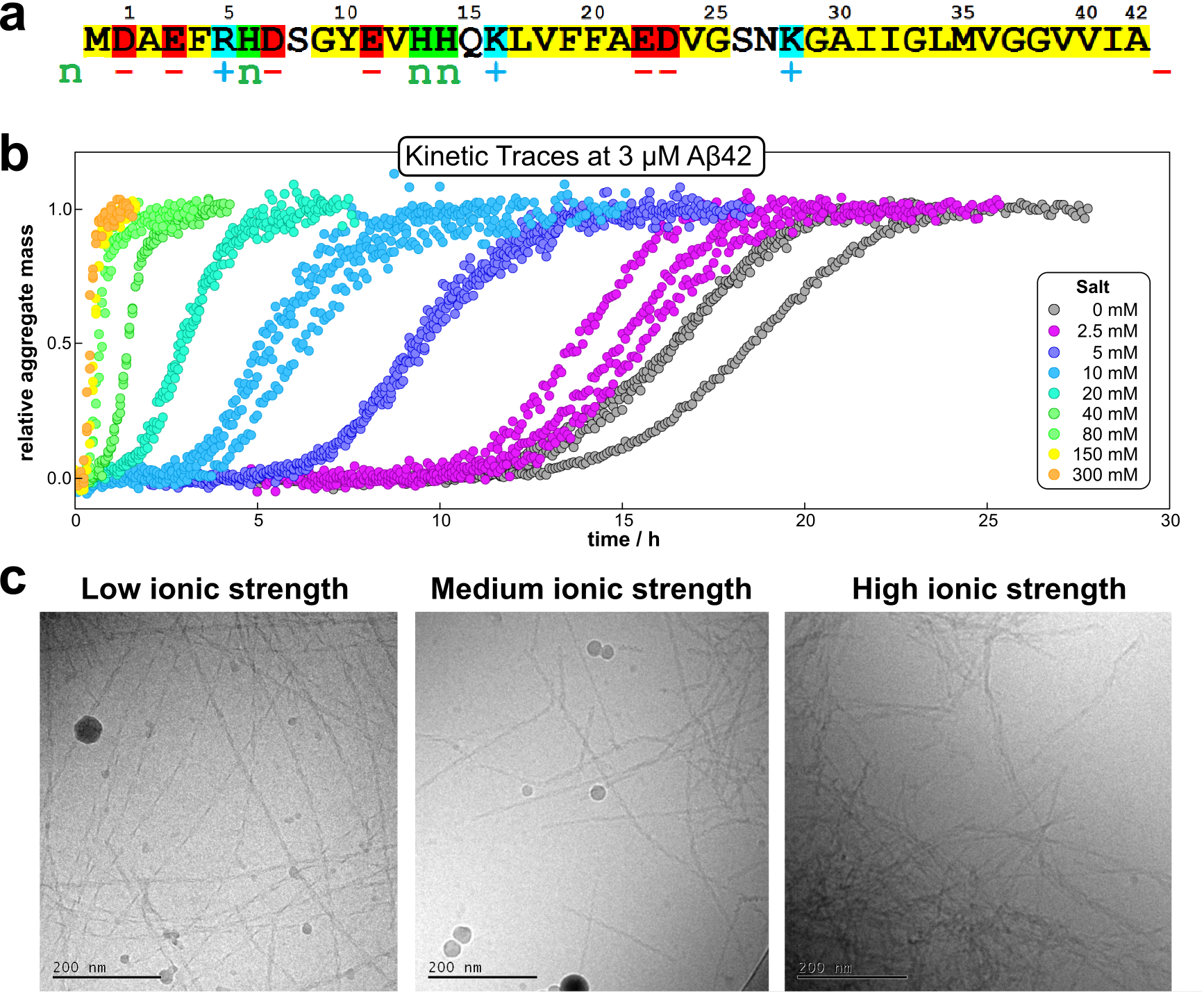}
	\caption{\textbf{Amino acid sequence and aggregation data for A$\beta$42.}  \textbf{a} Hydrophobic residues are shown in yellow, groups that are positively charged, negatively charged or having  pK value close to neutral pH are shown in blue (+), red (-) and green (n),  respectively. \textbf{b} The kinetic data for the aggregation of 3 $\mu$M A$\beta$42 in the presence of varying concentrations of salt, measured by ThT fluorescence; three replicates at each salt concentration are shown. \textbf{c} Cryo TEM images at a low (29 mM), intermediate (57 mM) and high (329 mM) ionic strength. The fibrils were obtained by aggregation of solutions with a monomer concentration of 10 $\mu$M. (Note: the large circular objects are particles of ice.) }
		\label{fig:sequence_and_TEM}
\end{figure}

Monomeric A$\beta$42 has a net charge of between -3 and -4 at pH 8.0 where the C-terminus and Asp, Glu, Lys and Arg side chains are mostly ionized and the His side chains and the N-terminus may be partly protonated (Fig.~\ref{fig:sequence_and_TEM}a)\cite{Klement2007,Betts2008}. Due to interactions of the charged groups, the specific value of the net charge is likely to change upon
the alteration of conformation and close packing associated with assembly of the peptides into aggregates\cite{Lund2005, silva2006}. The number of hydrophobic residues in the C-terminal segment of A$\beta$42 (Fig.~\ref{fig:sequence_and_TEM}a) gives this peptide a high aggregation propensity, despite the strong electrostatic repulsion between individual monomers.

In the present study the aggregation kinetics of A$\beta$42 at monomer concentrations ranging form 0.55 to 7 $\mu$M, and concentrations of up to 300 mM of added NaCl, in 4 mM phosphate buffer (giving an ionic strength of $\approx$12 mM in the absence of added NaCl), were recorded in triplicate repeats by monitoring thioflavin T (ThT) fluorescence (Fig.~\ref{fig:sequence_and_TEM}b). Fibrillar structures of similar morphology were found to be formed at all monomer and salt concentrations as monitored by cryo electron microscopy (TEM) (Fig.~\ref{fig:sequence_and_TEM}c), however, fibrils are packed more densely at higher salt concentrations. The entire set of experiments was repeated with a fresh batch of purified peptide, at salt concentrations in the same range, yielding similar results, which are shown in the SI, Fig.~S4%\ref{fig:kpkn_set2}
, S5%\ref{fig:rates_parallel}
 and S6%\ref{fig:rates_saturation}
. In order to analyse such a large body of kinetic data with a complex underlying mechanism we first set out to obtain general constraints on possible mechanisms by considering the qualitative features of the data.

\subsection*{Half Times and Scaling}
\begin{figure}[h!]
	\centering
		\includegraphics[width=\columnwidth]{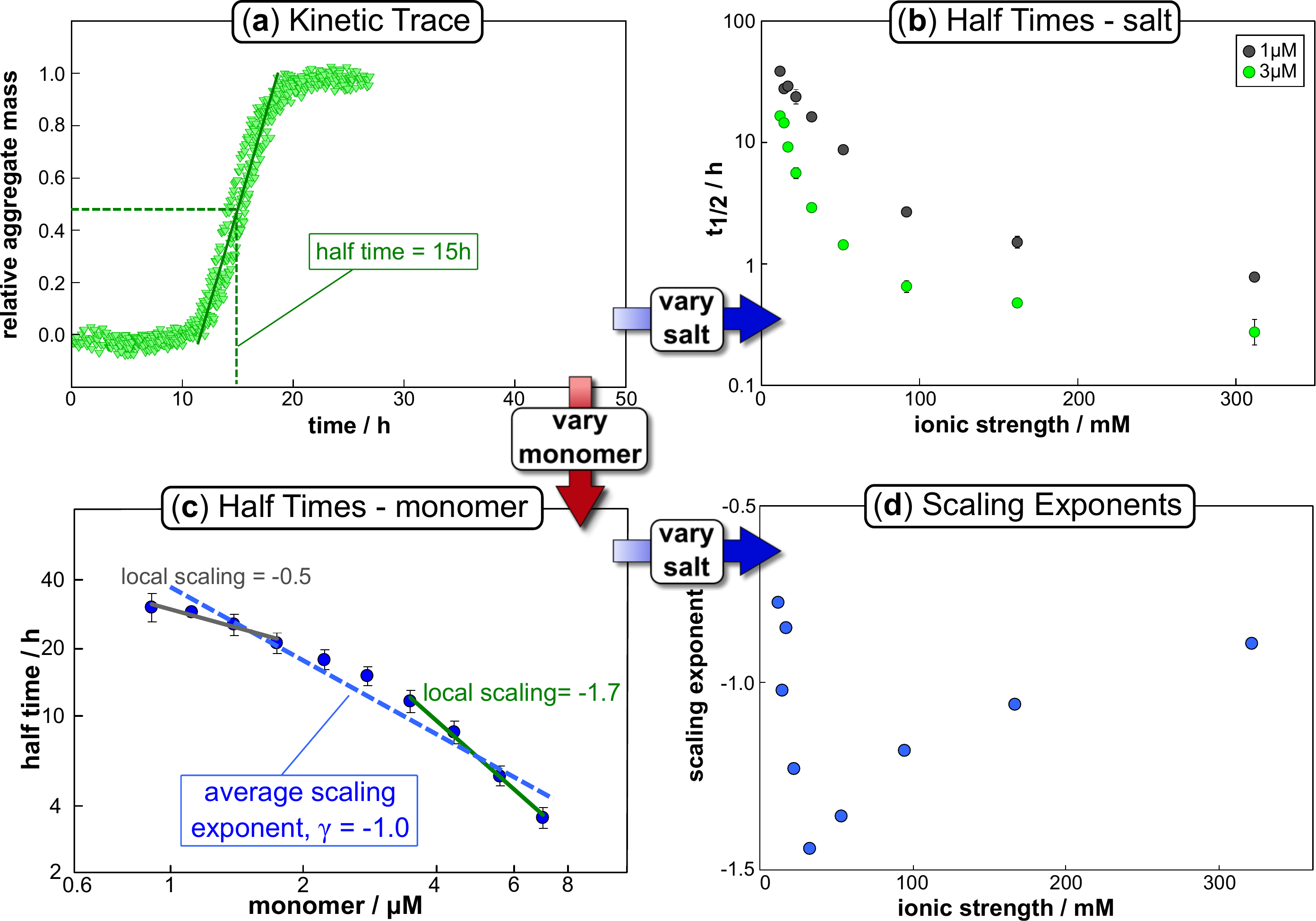}
	\caption{\textbf{Half time dependence on monomer and salt concentration.} The top left panel (a) shows a typical kinetic trace, in this case at a free monomer concentration of 3 $\mu$M and an ionic strength of 14.5 mM. The half time of aggregation can easily be extracted from such traces and plots of its variation with varying salt or monomer concentration are shown in the top right (b) and bottom left panels (c) respectively. The average slope of the double logarithmic plot of half time against monomer concentration gives the scaling exponent,$\gamma$, whose variation with ionic strength is plotted in the bottom right panel (d). Note the curvature in (c), indicative of a scaling exponent that depends on the monomer concentration.}
		\label{fig:halftimes_salt}
\end{figure}

The half time of the aggregation process is defined as the time by which half the final aggregate concentration has formed. In the first instance, the value of the half time is a guide to the aggregation propensity of a given system. For a charged peptide such as A$\beta$42, the aggregation propensity is expected to increase as the electrostatic repulsion between peptides becomes screened with increasing ionic strength. Indeed this accelerating effect of salt on the overall reaction rate was observed at all peptide concentrations examined in the present study, as is evident both from the kinetic curves in Fig.~\ref{fig:sequence_and_TEM}b and from the monotonic decrease in half times with increasing ionic strength, Fig.~\ref{fig:halftimes_salt}b. Similar results have been previously reported also for A$\beta$40 studied at a single peptide concentration \cite{Klement2007}.

To go beyond this very qualitative result and obtain mechanistic information, we consider the dependence of the half time $t_{1/2}$ on the monomer concentration, described by the scaling exponent $\gamma$ where $t_{1/2}\approx m_0^{\gamma}$ and $m_0$ is the monomer concentration at the beginning of an aggregation reaction. 
As outlined in Fig.~\ref{fig:halftimes_salt}a, the half time was determined for each kinetic curve at each monomer and salt concentration. The variation of the half time with varying salt concentration, at the same monomer concentration, shows the increase in aggregation propensity with increasing ionic strength, Fig.~\ref{fig:halftimes_salt}b. By contrast, from the variation of the half time with monomer concentration at the same salt concentration, the scaling exponent\cite{Cohen2011a} can be extracted, Fig.~\ref{fig:halftimes_salt}c, giving one scaling exponent at each salt concentration. 
We observed that the average scaling exponent has a biphasic dependence on the ionic strength, Fig.~\ref{fig:halftimes_salt}d; at low ionic strengths the half time scales with $\gamma=$-0.7. As the ionic strength increases, the average scaling exponent reaches a minimum of -1.4 at an ionic strength of around 32 mM and then increases again to -0.6 at an ionic strength of 312 mM, the highest value used here. 
Therefore, as the ionic strength is increased, the half times decrease monotonically for all monomer concentrations, but the dependence of the aggregation rate on the monomer concentration, given by the magnitude of the scaling, is largest at intermediate ionic strengths of circa 30 mM.

In addition we observe that at some salt concentrations there are variations in the scaling exponent as the monomer concentration changes, resulting in curvature of the double logarithmic plots of half time versus monomer concentration: At low ionic strengths the scaling exponent increases significantly in magnitude (i.e. the monomer dependence of the reaction increases) as the monomer concentration is increased, for example from $\gamma=-0.5$ to $\gamma=-1.7$ at an ionic strength of 14.5 mM, Fig.~\ref{fig:halftimes_salt}c. The opposite curvature is evident at some higher ionic strengths where the scaling exponent decreases in magnitude (i.e. the monomer dependence of the reaction decreases, Fig.~\ref{fig:mech_salt}) as the monomer concentration increases. 

In summary three distinct features emerge from this half time analysis: (1) The half times decrease with increasing ionic strength (as expected due to shielding of charge repulsion between peptides). (2) The half time scaling with monomer concentration is strongest at an ionic strength of approximately 30 mM and weaker at both lower and higher ionic strengths. (3) At some ionic strengths, the scaling exponent depends on the monomer concentration, i.e. there is curvature in double logarithmic plots of half time versus monomer concentration.

\begin{figure}[h!]
	\centering
		\includegraphics[width=0.5\columnwidth]{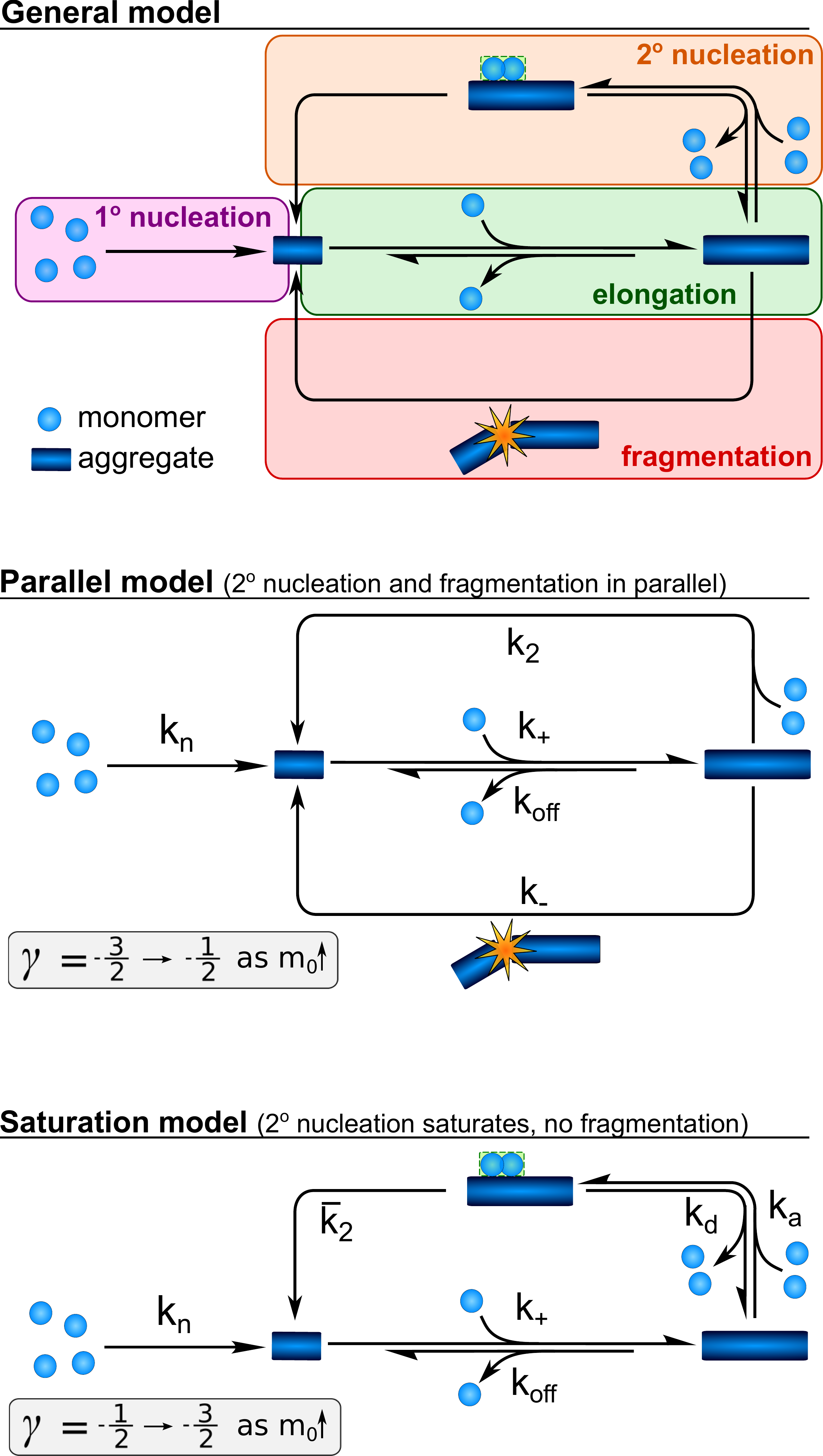}
	\caption{\textbf{A schematic depiction of the reaction network for the aggregation of A$\beta$42.} The general network (top), the special case when both fragmentation and unsaturated secondary nucleation are important (centre), resulting in a parallel network, and the special case where saturation of secondary nucleation is important (bottom) but fragmentation is negligible, resulting in a serial network. The ranges of scaling exponents covered by these models as the monomer concentration increases are given below each model. The rate constants given denote primary nucleation ($k_n$), elongation ($k_+$), depolymerisation ($k_{\mathrm{off}}$), fragmentation ($k_-$), single-step secondary nucleation ($k_2$), and the rates associated with multi-step secondary nucleation: the monomer attachment ($k_a$) and detachment rate constants ($k_d$), and the nucleus conversion/detachment rate constant ($\bar{k}_2$).}
		\label{fig:comic}
\end{figure}

\subsection*{Development of an aggregation model}
We next set out to develop a unifying model that reproduces these half time features and fits the time evolution of aggregate mass at all monomer and salt concentrations. Such a model will allow a determination of the underlying reaction network of the aggregation reaction and show how the dominant pathways through this network are shifted by changes in ionic strength.
In order to derive a quantitative description of the kinetics of an aggregating system\cite{ Ferrone1985b,Collins2004}, we have used a master equation approach\cite{Knowles2009,Cohen2011a, Meisl2016} that aims to classify all the processes that contribute significantly to the aggregation reaction by their mathematical dependence on the monomer concentration, the overall fibril number and the fibril mass concentrations. 

The elongation of the ends of growth competent fibril by the addition of monomers dominates the formation of fibril mass. In studies of the aggregation of other proteins, this elongation step has been found to saturate\cite{Esler2000a}, in our system, however, no saturation effects are observed (see SI Fig.~S8%\ref{fig:init_grad}
) hence elongation is modelled as a single step reaction. 

Whereas the production of fibril mass is dominated by a single process, three classes of processes are responsible for the formation of new aggregates and thereby new fibril ends (see Fig.~\ref{fig:comic}):  
(1) Processes that depend only on the monomer concentration, $m(t)$, (e.g. homogeneous nucleation of monomers in solution) which form new aggregates with rate constant $k_n$ and are of reaction order $n_c$ in monomer concentration (referred to as primary nucleation). 
(2) Processes that depend on the monomer concentration and the fibril mass (e.g. heterogeneous nucleation on the fibril surface) produce new aggregates with rate constant $k_2$  and are of order $n_2$ in monomer (referred to as secondary nucleation).
(3) Processes that depend only on the fibril mass (e.g. breakage of fibrils) and create new free ends with rate constant $k_-$ (referred to as fragmentation). The latter two types of event are referred to as secondary processes as they involve the catalysis of the formation of new aggregates by existing aggregates.

Whilst primary nucleation and fragmentation may be assumed to be single step processes, secondary nucleation consists of multiple steps in series, as we have established previously\cite{Meisl2014}. %, ***Saric2015}. 
The first step of secondary nucleation is a monomer-dependent attachment step in which free monomers interact with the fibril surface and the second step is a monomer-independent detachment of a newly formed nucleus. The overall process can be dominated by either step, depending on the conditions, and the concentration at which the second step becomes rate limiting, i.e. the concentration at which the fibril is fully covered in bound monomers, is determined by the constant $K_M$. By combining all these processes the full reaction network of aggregation is obtained, as displayed at the top of Fig.~\ref{fig:comic}.

Experiments in which the aggregation was monitored following the addition of monomeric A$\beta$42 to preformed fibrils confirmed that the production of new aggregates in the present system is dominated by secondary processes, as shown in detail in the SI (Fig.~S7%\ref{fig:low_seed}
). Under these conditions fragmentation and secondary nucleation produce significantly more new aggregates than primary nucleation even at very low aggregate concentrations\cite{Cohen2013}.
In order to simplify the fitting and analysis we have considered two special cases of this general reaction network, differing in their mechanism of the secondary process: In the first case both fragmentation and secondary nucleation, which act in parallel, may be significant, but we assume that there is no significant saturation of secondary nucleation (i.e. $K_M\gg m(t)^{n_2}$). We will refer to this case as the \textbf{parallel model} and  the corresponding scheme is shown in Fig.~\ref{fig:comic}.  
The second case allows for saturation of secondary nucleation, but assumes that there is no significant contribution from fragmentation (i.e. $k_2m(t)^{n_2}\gg k_-$). We will refer to this case as the \textbf{saturation model} and the corresponding scheme is shown in Fig.~\ref{fig:comic}c.
The detailed solutions to the kinetics of these models (see Methods and SI) can be used to estimate the expected range of scaling exponents which are shown below each model in Fig.~\ref{fig:comic}.

\begin{figure}
	\centering
\includegraphics[width=0.9\columnwidth]{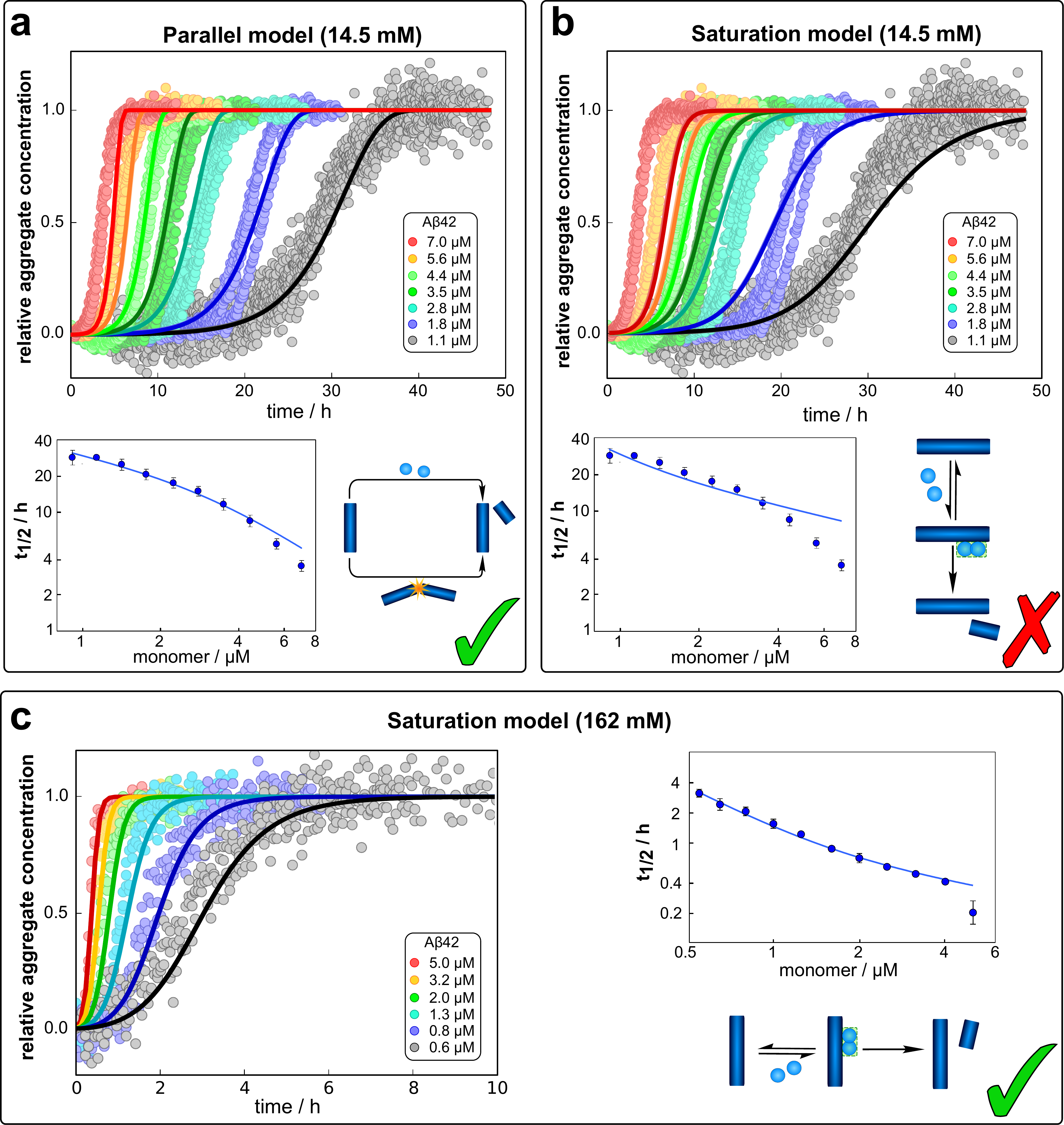}
\caption{\textbf{Global fits of the aggregation curves.} \textbf{a and b} The top panel shows a global fit of equations \ref{eq:M_par} (a) and \ref{eq:M_sat} (b) with three free parameters to the entire kinetic traces; shown below is a fit of just the half time behaviour.  The fit of the parallel model (a) is significantly better than that of the saturation model (b), the mean squared error of the parallel model being half of that of the saturation model. Note that the saturation model fails to reproduce the correct half times, as it cannot produce a decrease in scaling exponent with increasing monomer concentration. The saturation mechanism hence cannot explain the data at low ionic strengths. \textbf{c} A fit of the saturation model at high ionic strengths shows good agreement with the experimental data. For clarity not all sampled concentrations are displayed in the plots of the full timecourses; the fits were, however, performed on the entire dataset.  (See the SI, Fig.~S1%\ref{fig:2mM_150mM_fullfits}
, for plots of the full dataset and an explanation of the deviation of the parallel model at high monomer concentrations.)}
\label{fig:2mM_fit}
\end{figure}

\subsection*{Kinetic Analysis}
We fitted both models to the experimental data, obtained as described above, simultaneously for all monomer concentrations at each ionic strength\cite{Meisl2016}. The parallel model fits the data well at low and intermediate ionic strengths (see e.g. Fig.~\ref{fig:2mM_fit}a), but becomes invalid at high ionic strengths (see SI, Fig.~S5%\ref{fig:rates_parallel}
). The saturation model fits the data at intermediate and high ionic strengths (Fig.~\ref{fig:2mM_fit}c), but fails to reproduce the data at low ionic strengths (see Fig.~\ref{fig:2mM_fit}b).
\begin{figure*}
	\centering
\includegraphics[width=0.9\textwidth]{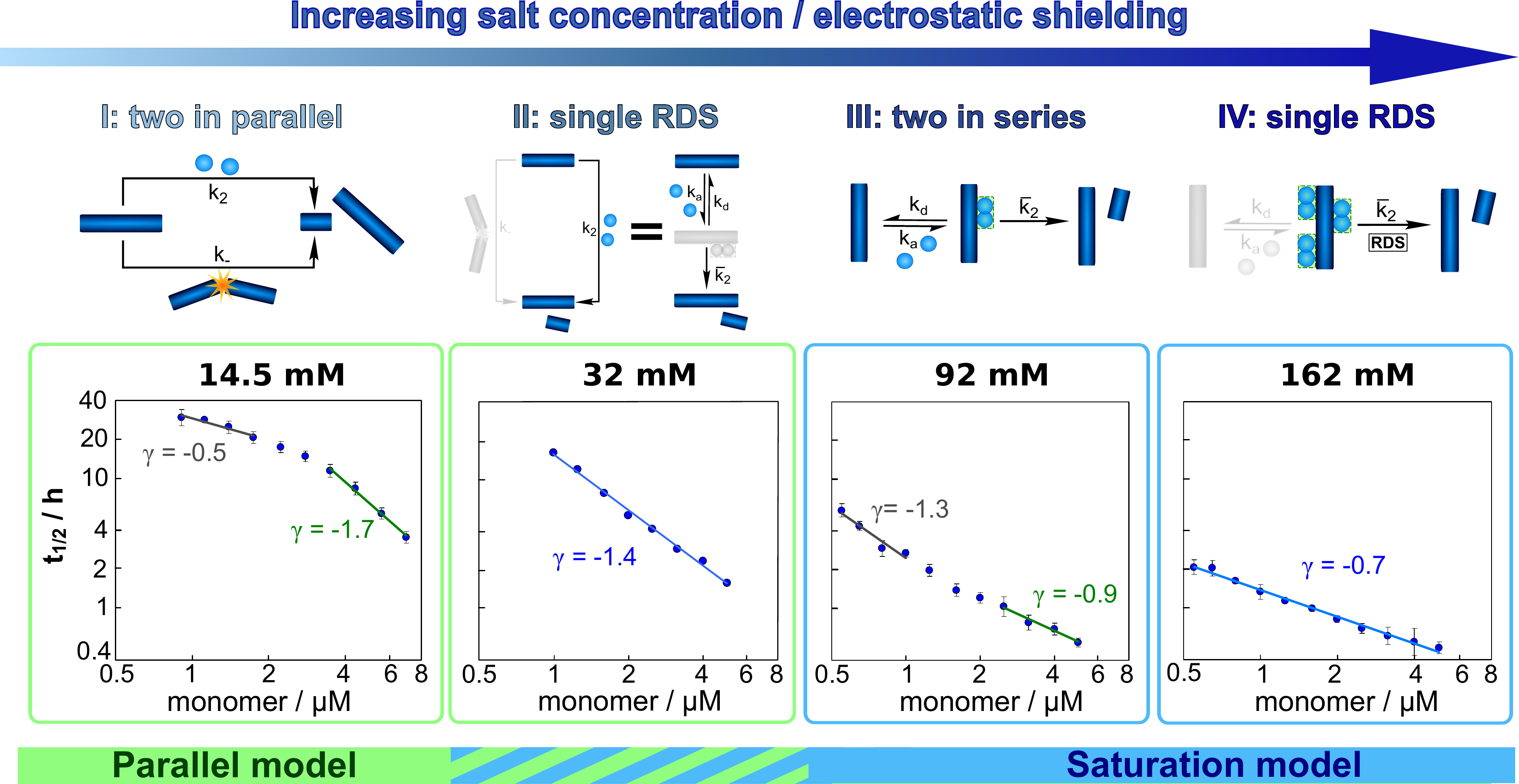}
\caption{\textbf{Dependence of the mechanism on ionic strength.} Double logarithmic plots of half time versus initial monomer concentration for representative salt concentrations, one from each of the four regimes detailed in the main text. Above each plot a schematic of the secondary process dominating the aggregation network is displayed. The differing slopes and curvature can be explained by considering which secondary mechanism dominates the generation of new fibrils. From left to right: At an ionic strength of 14.5 mM, a combination of fragmentation and secondary nucleation, with fragmentation dominating at low monomer concentrations and secondary nucleation at high monomer concentrations results in negative curvature. At an ionic strength of 32 mM, non-saturated secondary nucleation alone, i.e. a single rate determining step (RDS), describes the data over the entire monomer range. At an ionic strength of 92 mM, secondary nucleation saturates as the monomer concentration increases, giving positive curvature. At an ionic strength of 162 mM, secondary nucleation is fully saturated over the entire monomer concentration range, again requiring only a single RDS. }
\label{fig:mech_salt}
\end{figure*}

These results paint a clear picture of how the system goes through four distinct regimes, each defined by which mechanism dominates the production of new fibrils as the ionic strength increases (Fig.~\ref{fig:mech_salt}):

(I) At low ionic strengths (12 - 20 mM) the number of new growth competent ends produced by secondary nucleation and fragmentation is comparable (Fig.~\ref{fig:mech_salt}, 14.5 mM ionic strength): At low monomer concentrations fragmentation produces the largest number of new aggregates, resulting in a shallow dependence of the half time on monomer concentration, i.e. a scaling exponent of between -0.5 and -1. At high monomer concentrations secondary nucleation becomes the main source of new aggregates, resulting in a steeper dependence of the half time on monomer concentration with a scaling exponent of around -1.5. The average scaling exponent is circa -1.0.

(II) At intermediate ionic strengths (22 - 62 mM) fragmentation is negligible and secondary nucleation is not yet saturated: The monomer dependence of the half times is the same at all monomer concentrations, and the points in the double logarithmic plots of half time versus peptide concentration in Fig.~\ref{fig:mech_salt} at an ionic strength of 32 mM fall on a straight line. In this region both the parallel and the saturation models are equally valid. The average scaling exponent is close to its minimum value of -1.44. 

(III) At ionic strengths beyond values where the average scaling exponent is at a minimum (circa 92 mM) secondary nucleation starts to saturate: At low monomer concentrations it is barely saturated, giving a scaling exponent of circa -1.5, whereas at high monomer concentrations saturation starts to become significant, giving a scaling exponent of circa -1.0. Some curvature is evident in the half time plots (92 mM ionic strength Fig.~\ref{fig:mech_salt}).

(IV) At high ionic strengths (162 mM and above) secondary nucleation is fully saturated at all monomer concentrations investigated here. Repulsion between A$\beta$42 monomers is effectively screened and monomers have a high affinity to the fibril surface. Detachment of newly formed nuclei becomes rate limiting during secondary nucleation. The monomer dependence of the half times is constant with monomer concentration, the half times in Fig.~\ref{fig:mech_salt}, at an ionic strength of 162 mM, can be seen to lie on a straight line.

\subsection*{Effect of Ionic Strength on Individual Microscopic Processes}
The global fitting of the aggregation curves above provides an explanation of the observed changes in half times and scaling exponents: the differential effect of a change in ionic strength on the rates of the individual processes in the aggregation reaction leads to shifts in the dominant mechanism of aggregate multiplication. In order to rationalize this difference in susceptibility to electrostatic shielding we consider the variation of each rate constant with ionic strength.

\begin{figure}
	\centering
\includegraphics[width=\columnwidth]{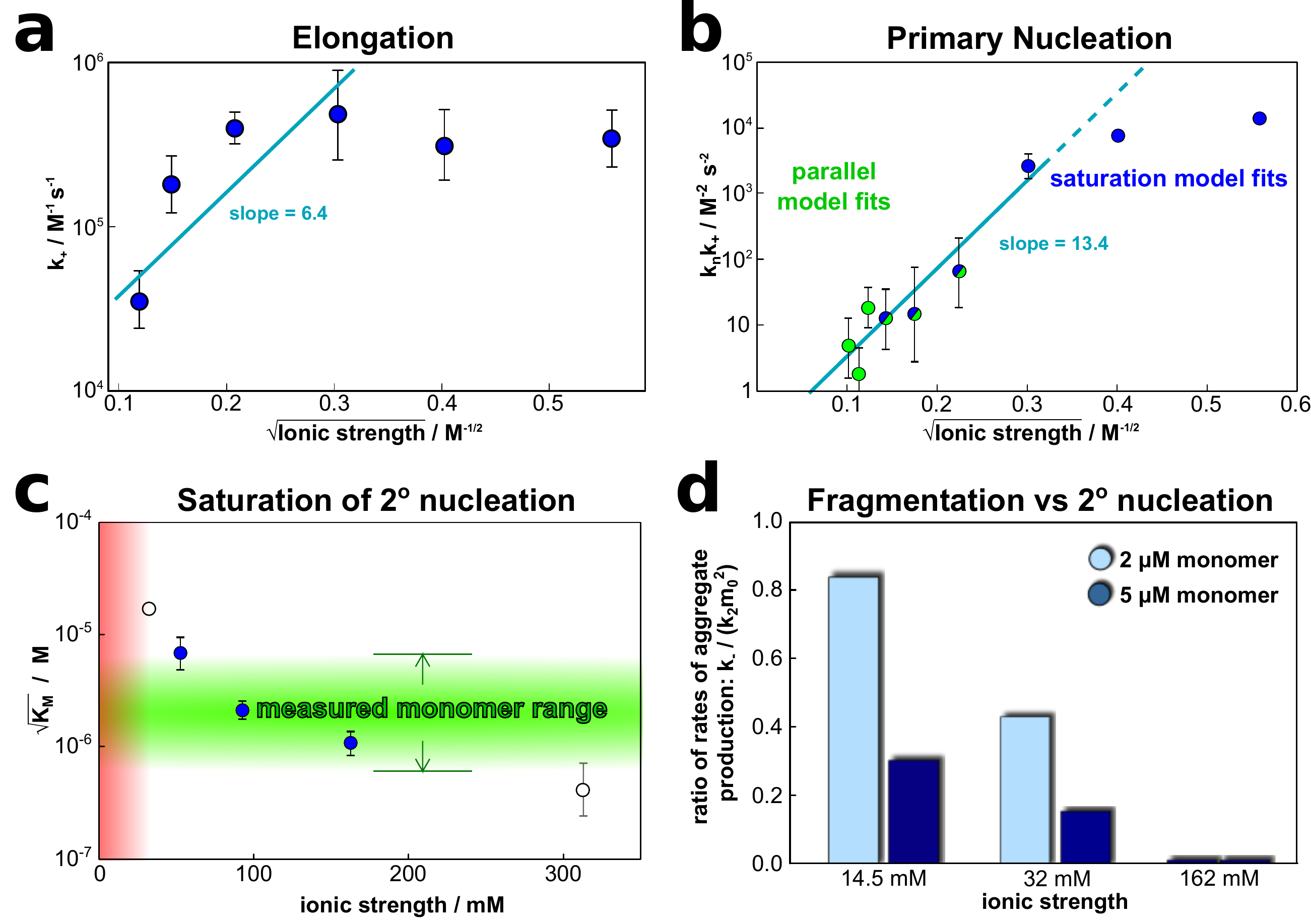}
\caption{\textbf{Effect of electrostatic screening on the microscopic rates.}  \textbf{a} The elongation rate constant as measured in strongly seeded experiments (blue dots, experimental details in SI). \textbf{b} The product of the elongation rate constant and the primary nucleation rate constant at different ionic strengths, obtained from global fits of the specific aggregation model that is valid at each ionic strength: Green dots from the parallel model (three lowest ionic strengths), blue dots from the saturation model (three highest ionic strengths); the green-blue dots are an average of the two models in the region of intermediate ionic strength where both models converge to the same limiting case (three intermediate ionic strengths). \textbf{c} The Michaelis constant, $K_M$, as obtained from global fits of the saturation model at high ionic strengths (the region where the saturation model is no longer valid is marked in red). As $\sqrt{K_M}$ gives the monomer concentration at which saturation effects become important, the region of monomer concentrations used in this study is marked in green. Values of $K_M$ outside the sampled region are likely to be less accurate, as the variation of $K_M$ in this region will have very little effect on the aggregation kinetics. The corresponding points are shown as empty circles.  \textbf{d} Ratio of the rates of production of new free ends from secondary nucleation and fragmentation at ionic strengths of 14.5 mM, 32 mM and 162 mM, at monomer concentrations of 2 $\mu$M (light blue) and 5 $\mu$M (dark blue). The ratio of fragmentation to secondary nucleation decreases both with increasing monomer concentration and increasing ionic strength. At ionic strengths of 162 mM and above, fragmentation is too slow to be measured.}
\label{fig:ratio_rates}
\end{figure}

In the plots in Fig.~\ref{fig:ratio_rates}a and b we show the logarithm of the various rate constant versus the square root of ionic strength (this latter value includes the contribution from the added salt as well as the buffer; in the absence of added salt the buffer alone (4 mM sodium phosphate, 40 uM EDTA, pH 8.0) results in an ionic strength of approximately 12 mM). In a simple Debye-Hueckel (DH) \cite{Debye1923,Bronsted1924, Schreiber2009} model of the effect of ionic strength, these points would be expected to lie on a straight line. However, DH is accurate only at low ionic strengths, which are not accessible experimentally due to the need for a buffer to control the pH and issues of irreproducibility which emerge in experiments with very slow aggregation rates and long lag times as observed at low ionic strengths. Straight line fits up to an ionic strength of 100 mM are shown in Fig.~\ref{fig:ratio_rates}, and although there is significant deviation from linearity, we can use the value of the slope to obtain on order of magnitude estimate of the charge of the species involved, as a consistency check. The slope corresponds approximately to a product of the valency of the reacting ions. In case of the elongation rate, Fig.~\ref{fig:ratio_rates}a, the slope is approximately 6, which is determined by the charge of a monomer and that of a free fibril end, consistent with a charge of -2 to -3 for each of these species. In the case of the combined elongation-nucleation rate constant, Fig.~\ref{fig:ratio_rates}b, the slope is approximately 13. If we assume the contribution from elongation is again approximately 6 (this is additive), the contribution from nucleation to the slope is then 7, which again is consistent with a charge of -2 to -3 for the monomers reacting during primary nucleation (details see SI). Hence, although DH theory is inadequate to describe fully the behaviour of charged macromolecules at the high ionic strengths studied here, the estimates it yields for the charges of the reacting species are entirely reasonable and hence consistent with an electrostatic effect.

\textit{Elongation rate constant:} The fitting of data obtained in the absence of the addition of any preformed fibrils yields the rate constants in the form of products, $k_+k_2$, $k_+k_n$ and $k_+k_-$, as the kinetics of such unseeded aggregation reactions depend only on these combinations rather than on the rate constants individually. An estimate for the elongation rate constant, $k_+$, can be obtained by performing experiments under strongly seeded conditions and measuring the initial increase in aggregate mass, which is determined only by the elongation of the seeds and not affected by nucleation processes. In order then to extract the value for the elongation rate constant, the number of seed fibrils needs to be determined. To this end, in the present study several TEM measurements were performed in order to obtain an estimate of the average length of the fibrils. The fibril lengths obtained in this way are only approximate and hence the elongation rates are estimated to be accurate only to within an order of magnitude; the absolute value of $k_+$ should therefore be interpreted with caution. Its relative variation with ionic strength does, however, not suffer from such inaccuracies (details see SI). The elongation rate is found to increase by approximately one order of magnitude as the ionic strength is varied between 12 and 52 mM, Fig.~\ref{fig:ratio_rates}a.

\textit{Nucleation rate constants:} The product of the primary nucleation rate constant and the elongation rate constant is the only quantity common to both models and hence can be obtained for the entire range of salt concentrations from the global fits. In the region of intermediate ionic strength (22 mM to 62 mM) the two models converge to the same limit (single step secondary nucleation without fragmentation) and hence 
in Fig.~\ref{fig:ratio_rates}b, in the intermediate region, the average of the fits is shown. The combined primary nucleation/elongation rate constant is found to increase by four orders of magnitude from the lowest to the highest ionic strength, which implies an increase of the primary nucleation rate constant by two to three orders of magnitude, and hence that primary nucleation is affected more strongly than elongation by the electrostatic shielding. This may be due to the fact that during primary nucleation a larger number of charged species come together to form the nucleus.
The secondary nucleation rate constant shows an increase similar to that of the primary nucleation rate constant, whereas the fragmentation rate constant remains approximately constant (plots see Fig.~\ref{fig:ratio_rates}b and Fig.~S5%\ref{fig:rates_parallel}
, Fig.~S6%\ref{fig:rates_saturation}
). The increase in the ratio of nucleation rate to elongation rate is also in agreement with the decrease in fibril length observed in the TEM measurements (see SI, Fig.~S9%\ref{fig:TEM_lengths}
).

\textit{Saturation concentration:} From fits of the saturation model, the monomer concentration at which saturation of secondary nucleation occurs is found to decrease with increasing ionic strength. $\sqrt{K_M}$ gives the monomer concentration at which saturation effects become important: if $\sqrt{K_M}$ is above the region of monomer concentrations that are sampled in the experiment we expect the system to be unsaturated, if $\sqrt{K_M}$ is within the region of monomer concentrations that are sampled we expect the system to start displaying saturation as the monomer concentration increases, and finally if the region of monomer concentrations that are sampled is above $\sqrt{K_M}$ we expect the system to be fully saturated at all sampled monomer concentrations. The results show that $\sqrt{K_M}$ decreases with increasing ionic strength, i.e. the system will begin to saturate at lower monomer concentrations the higher the ionic strength (Fig.~\ref{fig:ratio_rates}c).
This observation can be rationalised by considering the fact that the monomer concentration dependent step in secondary nucleation involves an association reaction between negatively charged monomers and negatively charged fibrils, which will be slower at lower ionic strengths. By contrast the monomer independent detachment step is dominated by short range interactions and rearrangements, which is therefore likely to be less affected by the screening of electrostatic interactions. At high ionic strengths the first, monomer-dependent, step becomes very fast, whereas the monomer-independent step proceeds at a rate similar to that at low ionic strength, and therefore it becomes rate limiting, leading to the observed saturation of secondary nucleation. 

\textit{Dominant secondary process:} Which of the two secondary processes dominates the multiplication of aggregates is determined by both the ionic strength and the monomer concentration. To illustrate this point, we have considered the relative number of new aggregates produced by secondary nucleation and by fragmentation, at three different ionic strengths and two monomer concentrations (Fig.~\ref{fig:ratio_rates}d).  At an ionic strength of 14.5 mM and low monomer concentrations, the ratio of fragmentation to secondary nucleation is close to 1, and the two processes both contribute to the kinetics. At higher monomer concentrations, secondary nucleation is faster, as its rate increases with monomer concentration whereas fragmentation is independent of the monomer concentration. As the  ionic strength is increased to 32 mM, the rate of secondary nucleation increases significantly, whereas the fragmentation rate does not match this increase, making the contribution of fragmentation to the aggregation reaction insignificant even at low monomer concentrations. At even higher ionic strengths, the fragmentation rate is too low to be detected in the fitting.

Although the aggregation behaviour of A$\beta$42 is evidently complex over the entire salt and monomer concentration ranges, it can be rationalised completely by considering the effect of ionic strength in terms of an increase of the association rates of the various microscopic processes. The more charged species associate in a given process, the more it is affected by a change in ionic strength. The fact that this simple explanation is sufficient to account for such a large range of complex behaviours strongly supports the minimal mechanistic description of the aggregation process developed here. 

\begin{table*}
\centering
\caption{ \textit{\textbf{Reaction network to unify variants of A$\beta$. }}}
\small
\begin{tabularx}{\textwidth}{ |l |l|l| X| X| } % use 'Y' automatic columnwidth with linebreak
\hline
&& \textbf{proposed}&\textbf{mechanistic} &\\
\textbf{system} & \textbf{scaling} & \textbf{mechanism} & \textbf{analogue} & \textbf{rationalisation}\\
 \hline  
A$\beta$40\cite{Meisl2014} & -1.2 to -0.5 & saturating 2$^o$ nuc & medium to high ionic strength & decrease in detachment rate during 2$^o$ nuc \\ \hline  
A$\beta$42, shaking\cite{Cohen2013} & -0.6 & fragmentation & low ionic strength, low monomer concentration & increase in fragmentation due to shear forces \\    \hline 
A$\beta$42 A2V\cite{Meisl2016a} & -1.5 to -0.5 & saturating 2$^o$ nuc & medium to high ionic strength & increased attachment during 2$^o$ nuc, due to increased hydrophobicity \\    \hline
A$\beta$42 pH7.4\cite{Meisl2016a} & -1.5 to -0.5 & saturating 2$^o$ nuc & medium to high ionic strength & increased attachment during 2$^o$ nuc, due to decreased charge \\   \hline
A$\beta$42 N-term ext\cite{Szczepankiewicz2015} & approx -1.3 & unsat 2$^o$ nuc & medium ionic strength &  unchanged from A$\beta$42\\   \hline
A$\beta$42 E22G\cite{Bolognesi2014} & -0.5 & fully sat 2$^o$ nuc & high ionic strength & increased attachment during 2$^o$ nuc, due to decreased charge \\     
 \hline
\end{tabularx}
\label{table:comparison}
\normalsize
\end{table*}

\section*{Concluding remark}
In this work we have shown that electrostatic screening can be used to modulate the relative importance of different microscopic processes and thereby to alter the specific pathway that dominates the conversion of soluble peptides to their amyloid forms.  This approach has allowed us to tackle a complex reaction network and to establish a mechanistic framework that accounts fully for the aggregation behaviour of A$\beta$42 over a wide range of solution conditions and monomer concentrations.
As the individual rate constants are altered by the increasing electrostatic shielding, modifying the interplay and competition of the different microscopic processes, four distinct types of macroscopic behaviour emerge. The rate of fragmentation is found to be largely unaffected by variations in ionic strength and only contributes notably at low ionic strengths and monomer concentrations. By contrast, the rate constants for primary and secondary nucleation each increase by approximately two orders of magnitude and the rate constant of elongation increases by circa one order of magnitude, upon increasing the ionic strength from 12 mM to 312 mM. The magnitude of the effect correlates with the number of charged species that come together in the respective processes of nucleation (two or more), elongation (two) and fragmentation (none). 
These findings show that the modulation of electrostatic interactions has profound mechanistic effects beyond a simple overall increase in aggregation propensity, and due to its differential effect on the individual rates it represents a means to sample an extended reaction network.

The general reaction network we obtained here through this sampling of a large proportion of the space of possible aggregation mechanisms serves as a unified framework for describing the aggregation behaviour of the variants of the A$\beta$ peptide. Indeed many variants of  A$\beta$ under a range of conditions aggregate via a pathway that is part of this general reaction network, summarised in table \ref{table:comparison}. Generally we find that effects that increase the fibril coverage during secondary nucleation or increase the rate of attachment of monomers to fibrils relative to the detachment of newly formed nuclei lead to an aggregation mechanism that is dominated by a saturating, or saturated, secondary nucleation mechanism, a process often associated with toxicity\cite{Walsh2002, Jan2011}. This is the case for the aggregation of A$\beta$42 at a slightly lower pH of 7.4\cite{Meisl2016a}, as well as the aggregation of the mutants A2V\cite{Meisl2016a} and E22G\cite{Bolognesi2014}. The reasons for increased fibril coverage in these cases are believed to be a decreased electrostatic repulsion as more residues become protonated at lower pH, a higher hydrophobicity and hence stronger binding of monomers to fibrils for A2V, and again a lowering of the electrostatic repulsion due to the loss of a charged sidegroup for E22G. This behaviour is analogous to the increased fibril affinity of monomers we observed here at high ionic strengths. We previously found that the other major variant of the A$\beta$ peptide, A$\beta$40, aggregates via a saturating secondary nucleation mechanism, however the overall rates are lower than in the case of A$\beta$42, possibly due to a decreased nucleus formation / detachment step, rather than an increased binding, during secondary nucleation.  By contrast, N-terminally extended variants of  A$\beta$42 aggregate via an unsaturated secondary nucleation mechanism, which is likely to be due to the reduced number of reactive encounters of fibrils and monomers and therefore a saturation of the fibril surface is not reached at the studied monomer concentrations\cite {Szczepankiewicz2015}. 
Finally our extended reaction network also connects the aggregation behaviour of A$\beta$42 under strong agitation\cite{Cohen2013}, with the behaviour under quiescent conditions: the shear forces induced by shaking significantly increase the rate of fragmentation and the mechanism shifts to one dominated by fragmentation. An analogous effect can be produced by significantly lowering the rate of the other nucleation processes, as we have observed here at low salt concentrations, allowing fragmentation to become kinetically visible. 

The analysis detailed here, therefore, serves as a basis for interpreting how changes in solution conditions or peptide sequence shift the dominant pathways in the reaction network. It provides a continuum of mechanisms connecting the varied behaviour of different systems. In particular it allows the determination of the extent to which any alterations of the peptide sequence or the solution conditions correlate with those produced by altered electrostatic screening, thereby providing key insights into the physical origin of any observed mechanistic differences. 

\section*{Materials and methods}

\subsection*{A$\beta$42 Expression and Purification} A$\beta$(M1-42), of sequence shown in Fig.~\ref{fig:sequence_and_TEM}, here referred to as A$\beta$42, was expressed in E. coli from a synthetic gene and purified in batch format using ion exchange and size exclusion steps as described  in Walsh \textit{et. al}\cite{Walsh2009}. This results in highly pure monomeric peptide, which was divided in identical aliquots and stored as lyophilized peptide powder.

\subsection*{Samples for aggregation kinetics} 
Monomeric A$\beta$42 was isolated twice by gel filtration of aliquots of purified peptides just prior to preparing each of the experiments, to remove any traces of aggregates formed during freezing and thawing and to exchange the buffer. The peptide powder was dissolved in 6 M GuHCl at pH 8.0 and the monomer peak was collected in low-bind Eppendorf tubes (Axygene) on ice and the concentration was determined by absorbance at 280 nm using an extinction coefficient of 1440 l mol$^{-1}$ cm$^{-1}$. Gel filtration was performed in 20 mM sodium phosphate, 0.2 mM EDTA, pH 8 at high enough concentrations that the isolated monomeric peptide could be diluted 5-fold with water or NaCl and still contain monomer concentrations of up to 5-7 $\mu$M. The final solutions used in the aggregation assays contained 4 mM sodium phosphate buffer, pH 8.0, with 40 $\mu$M EDTA, 6 $\mu$M ThT and NaCl in the range between 0 to 300 mM. The monomer concentration was varied between 0.5 and 7 $\mu$M and the ThT concentration (6 $\mu$M) was chosen to be in the range that has been shown to provide a response that is linearly dependent on the total aggregate mass \cite{Cohen2013}. Here, this linearity was found to hold at all salt concentrations. At 0-10 mM NaCl the fluorescence intensity is higher the lower the salt concentration, but the response still varies linearly with monomer concentration for each salt concentration. All solutions used in gel filtration and in the preparation of sample series were extensively degassed.
The aggregation kinetics were studied at 37$^{\mathrm{o}}$C under quiescent condition by recording the ThT fluorescence intensity as a function of time using a plate reader (FluoStar Omega or Optima, BMG Labtech, Offenberg, Germany). The fluorescence was recorded in half-area 96-well PEG-coated black polystyrene plates with clear bottoms (Corning 3881, Massachusetts, USA) measuring from below with a 440 nm excitation filter and a 480 nm emission filter.

\subsection*{Theoretical Model}
The time evolution of aggregate mass, $M(t)$, for the parallel model is given by:

\begin{eqnarray}
M(t)&=&m_{\mathrm{tot}}+\mathrm{exp}\left[-\frac{k_+(4c\kappa\mathrm{cosh}(\kappa t)+4P_0\kappa^2\mathrm{sinh}(\kappa t))}{2\kappa^3}\right] \nonumber \\ 
&&\left((M_0-m_{\mathrm{tot}})e^{\frac{2k_+c}{\kappa^2}}\right)
\label{eq:M_par}
\end{eqnarray}
where
\begin{eqnarray}
a&=&k_2m_0^{n_2}+k_- \nonumber \\
c&=&k_nm_0^{n_c}+aM_0 \nonumber \\
\kappa&=&\sqrt{2k_+m_0(k_2m_0^{n_2}+k_-)}
\label{eq:}
\end{eqnarray}
where $m_{\mathrm{tot}}$ is the total protein concentration, $M_0$, $P_0$ and $m_0$ are the initial mass concentration of fibrils, number concentration of fibrils and monomer concentration, respectively. $k_+$, $k_-$, $k_n$ and $k_2$ are the rate constants of elongation, fragmentation, primary nucleation and secondary nucleation.

For the saturation model we obtain:

\begin{eqnarray}
\frac{M}{M_{\infty}}&=&1-\left(1-\frac{M_0}{M_{\infty}}\right)e^{-k_{\infty}t}\nonumber \\
&&\cdot \left(\frac{B_-+C_+e^{\kappa t}}{B_++C_+e^{\kappa t}}\cdot\frac{B_++C_+}{B_-+C_+}\right)^{\frac{k_{\infty}}{\kappa \bar{k}_{\infty}}}
\label{eq:M_sat}
\end{eqnarray}
where the definitions of the parameters are
\begin{eqnarray}
\kappa&=&\sqrt{2m_0k_+\frac{m_0^{n_2}k_2}{1+m_0^{n_2}/K_M}}\\
\lambda&=&\sqrt{2k_+k_nm_0^{n_c}}\\
C_{\pm}&=&\frac{k_+P_0}{\kappa}\pm \frac{k_+M_0}{2m_0k_+}\pm \frac{\lambda^2}{2\kappa^2}\\
k_{\infty}&=&2k_+P_{\infty}\\
\bar{k}_{\infty}&=&\sqrt{k_{\infty}^2-2C_+C_-\kappa^2}\\
B_{\pm}&=&\frac{k_{\infty}\pm\bar{k}_{\infty}}{2\kappa}
\end{eqnarray}
Again $m_0$ is the initial monomer concentration, $P_0$, $M_0$ and $P_{\infty}$, $M_{\infty}$ are the aggregate number and mass concentrations at the beginning of the reaction and at equilibrium, i.e. after completion of the aggregation reaction.
The details of the derivation of these models can be found in the SI.

Note that for unseeded  experiments (i.e. experiments starting from monomer alone, without preformed fibrils) the two limiting cases only involve three free parameters each ($k_+k_2$, $k_+k_n$ and $k_+k_-$ for the parallel model and $k_+k_2$, $k_+k_n$ and $K_M$ for the saturation model), which were found to be sufficient in the present study to produce high quality global fits to all monomer concentrations at any given salt concentration.
For the parallel model, the scaling exponents is given by:

\begin{equation}
\gamma_{\mathrm{comp}}= \frac{d\log(t_{1/2})}{d\log(m(0))}\approx       -\frac{1}{2}\left(\frac{n_2}{1+K/m(0)^{n_2}}+1\right)
\label{eq:scaling_frag_secnuc}
\end{equation}
where $K=k_-/k_2$. This expression interpolates between  $\gamma = -1/2$ and $\gamma = -(n_2+1)/2$ for the limits of low and high monomer concentration respectively, giving the negative curvature in the double logarithmic plots of the half time, as predicted from the qualitative argument used to derive the general constraints above.

For the saturation model, the scaling exponent is given by:
\begin{equation}
\gamma_{\mathrm{sat}}\approx       -\frac{1}{2}\left(\frac{n_2}{1+m(0)^{n_2}/K_M}+1\right)
\label{eq:scaling_secnuclim}
\end{equation}
where $K_M$ is the Michaelis constant. This expression interpolates between  $\gamma = -(n_2+1)/2$ and $\gamma = -1/2$  for the limits of low and high monomer concentration respectively, i.e. the reverse of the parallel model.

\subsection*{Fitting} The data were  normalised to give units of relative fibril mass concentration. The fits were performed using a basin-hopping algorithm\cite{Wales1997,Meisl2014} on the AmyloFit fitting platform\cite{Meisl2016}. There were three global (i.e. one value of each parameter for all monomer concentrations) fitting parameters at each salt concentration: $k_nk_+$,$k_2k_+$ and $k_-k_+$ for the parallel model, and $k_nk_+$,$k_2k_+$ and $K_M$ for the saturation model. The reaction orders of primary and secondary nucleation, $n_c$ and $n_2$, were both fixed to 2, in line with the values found for A$\beta$42 in previous work\cite{Cohen2013}. The effect of fitting the reaction orders, which were set to the value previously established at a single salt concentration\cite{Cohen2013}, is discussed in the SI.


\newpage

\begin{thebibliography}{10}

\bibitem{Lawrence2007}
Michael~S. Lawrence, Kevin~J. Phillips, and David~R. Liu.
\newblock Supercharging proteins can impart unusual resilience.
\newblock {\em Journal of the American Chemical Society}, 129(33):10110--10112,
  2007.

\bibitem{Kurnik2012}
Martin Kurnik, Linda Hedberg, Jens Danielsson, and Mikael Oliveberg.
\newblock Folding without charges.
\newblock {\em Proceedings of the National Academy of Sciences},
  109(15):5705--5710, 2012.

\bibitem{marti2000}
Daniel~N. Marti and Hans~Rudolf Bosshard.
\newblock Inverse electrostatic effect: Electrostatic repulsion in the unfolded
  state stabilizes a leucine zipper.
\newblock {\em Biochemistry}, 43(39):12436--12447, 2004.

\bibitem{lindman2006a}
Stina Lindman, Sara Linse, Frans A.~A. Mulder, and Ingemar Andre.
\newblock Electrostatic contributions to residue-specific protonation
  equilibria and proton binding capacitance for a small protein.
\newblock {\em Biochemistry}, 45(47):13993--14002, 2006.

\bibitem{silva2005}
Fernando L. B. da~Silva, Sara Linse, and Bo~J\"onsson.
\newblock Binding of charged ligands to macromolecules. anomalous salt
  dependence.
\newblock {\em The Journal of Physical Chemistry B}, 109(5):2007--2013, 2005.

\bibitem{Kesvatera1996}
Tonu Kesvatera, Bo~Jonsson, Eva Thulin, and Sara Linse.
\newblock Measurement and modelling of sequence-specific pkavalues of lysine
  residues in calbindin \{D9k\}.
\newblock {\em Journal of Molecular Biology}, 259(4):828 -- 839, 1996.

\bibitem{kesvatera1999}
T\"onu Kesvatera, Bo~J\"onsson, Eva Thulin, and Sara Linse.
\newblock Ionization behavior of acidic residues in calbindin d9k.
\newblock {\em Proteins: Structure, Function, and Bioinformatics},
  37(1):106--115, 1999.

\bibitem{Linse1988}
Sara Linse, Peter Brodin, Charlotta Johansson, Eva Thulin, Thomas Grundstrom,
  and Sture Forsen.
\newblock The role of protein surface charges in ion binding.
\newblock {\em Nature}, 335(6191):651--652, 1988.

\bibitem{Zhou2005}
Huan-Xiang Zhou.
\newblock How do biomolecular systems speed up and regulate rates?
\newblock {\em Physical Biology}, 2(3):R1, 2005.

\bibitem{Matousek2007}
William~M. Matousek, Barbara Ciani, Carolyn~A. Fitch, Bertrand~E.
  Garcia-Moreno, Richard~A. Kammerer, and Andrei~T. Alexandrescu.
\newblock Electrostatic contributions to the stability of the \{GCN4\} leucine
  zipper structure.
\newblock {\em Journal of Molecular Biology}, 374(1):206 -- 219, 2007.

\bibitem{Lindman2006b}
Stina Lindman, Wei-Feng Xue, Olga Szczepankiewicz, Mikael~C. Bauer, Hanna
  Nilsson, and Sara Linse.
\newblock Salting the charged surface: ph and salt dependence of protein g
  \{B1\} stability.
\newblock {\em Biophysical Journal}, 90(8):2911 -- 2921, 2006.

\bibitem{Xue2009}
Wei-Feng Xue, Andrew~L Hellewell, Walraj~S Gosal, Steve~W Homans, Eric~W
  Hewitt, and Sheena~E Radford.
\newblock Fibril fragmentation enhances amyloid cytotoxicity.
\newblock {\em J Biol Chem}, 284(49):34272--34282, Dec 2009.

\bibitem{Vendruscolo2007}
Michele Vendruscolo and Christopher~M. Dobson.
\newblock Chemical biology: More charges against aggregation.
\newblock {\em Nature}, (449):555, 2007.

\bibitem{Chiti2006}
Fabrizio Chiti and Christopher~M Dobson.
\newblock Protein misfolding, functional amyloid, and human disease.
\newblock {\em Annu Rev Biochem}, 75:333--366, 2006.

\bibitem{Knowles2014}
Tuomas P~J Knowles, Michele Vendruscolo, and Christopher~M. Dobson.
\newblock The amyloid state and its association with protein misfolding
  diseases.
\newblock {\em Nature Reviews Molecular Cell Biology}, 15:384--396, 2014.

\bibitem{Schreiber2009}
Gideon~Schreiber, Gilad~Haran, and Huan-Xiang Zhou.
\newblock Fundamental aspects of protein−protein association kinetics.
\newblock {\em Chemical Reviews}, 109(3):839--860, 2009.

\bibitem{Buell2013}
Alexander K. Buell, Peter Hung, Xavier Salvatella, Mark E. Welland,
  Christopher M. Dobson, and Tuomas P.J. Knowles.
\newblock Electrostatic effects in filamentous protein aggregation.
\newblock {\em Biophysical Journal}, 104:1116--1126, 2013.

\bibitem{Abelein2015}
Axel Abelein, Astrid Graslund, and Jens Danielsson.
\newblock Zinc as chaperone-mimicking agent for retardation of amyloid $\beta$
  peptide fibril formation.
\newblock {\em Proceedings of the National Academy of Sciences},
  112(17):5407--5412, 2015.

\bibitem{Knowles2009}
Tuomas P.~J. Knowles, Christopher~A. Waudby, Glyn~L. Devlin, Samuel I.~A. Cohen,
  Adriano Aguzzi, Michele Vendruscolo, Eugene~M Terentjev, Mark~E Welland, and
  Christopher~M Dobson.
\newblock An analytical solution to the kinetics of breakable filament
  assembly.
\newblock {\em Science}, 326(5959):1533--1537, Dec 2009.

\bibitem{Cohen2011a}
Samuel I.~A. Cohen, Michele Vendruscolo, Mark~E. Welland, Christopher~M. Dobson,
  Eugene~M Terentjev, and Tuomas P~J Knowles.
\newblock Nucleated polymerization with secondary pathways. i. time evolution
  of the principal moments.
\newblock {\em J Chem Phys}, 135(6):065105, Aug 2011.

\bibitem{Meisl2016}
Georg Meisl, Julius~B. Kirkegaard, Paolo Arosio, Thomas T.~C. Michaels, Michele
  Vendruscolo, Christopher~M. Dobson, Sara Linse, and Tuomas P.~J. Knowles.
\newblock Molecular mechanisms of protein aggregation from global fitting of
  kinetic models.
\newblock {\em Nature Protocols}, 11(2):252--272, 2016.

\bibitem{Klement2007}
Karolin Klement, Karin Wieligmann, Jessica Meinhardt, Peter Hortschansky,
  Walter Richter, and Marcus F{\"a}ndrich.
\newblock Effect of different salt ions on the propensity of aggregation and on
  the structure of {A}lzheimer's abeta(1-40) amyloid fibrils.
\newblock {\em J Mol Biol}, 373(5):1321--1333, Nov 2007.

\bibitem{Betts2008}
Vicki Betts, Malcolm~A. Leissring, Georgia Dolios, Rong Wang, Dennis~J. Selkoe,
  and Dominic~M. Walsh.
\newblock Aggregation and catabolism of disease-associated intra-a$\beta$
  mutations: reduced proteolysis of a$\beta$a21g by neprilysin.
\newblock {\em Neurobiology of Disease}, 31(3):442 -- 450, 2008.

\bibitem{Lund2005}
Mikael Lund and Bo~J\"onsson.
\newblock On the charge regulation of proteins.
\newblock {\em Biochemistry}, 44(15):5722--5727, 2005.

\bibitem{silva2006}
Fernando L.~B. da~Silva, Mikael Lund, Bo~J\"onsson, and Torbj\"orn Akesson.
\newblock On the complexation of proteins and polyelectrolytes.
\newblock {\em The Journal of Physical Chemistry B}, 110(9):4459--4464, 2006.

\bibitem{Ferrone1985b}
Frank~A. Ferrone, James~Hofrichter, and William~A. Eaton.
\newblock Kinetics of sickle hemoglobin polymerization. ii. a double nucleation
  mechanism.
\newblock {\em J Mol Biol}, 183(4):611--631, Jun 1985.

\bibitem{Collins2004}
Sean~R. Collins, Adam Douglass, Ronald~D. Vale, and Jonathan~S. Weissman.
\newblock Mechanism of prion propagation: amyloid growth occurs by monomer
  addition.
\newblock {\em PLoS Biol}, 2(10):e321, Oct 2004.

\bibitem{Esler2000a}
William~P. Esler,  Evelyn~R. Stimson, Joan~M. Jennings, Harry~V. Vinters, Joseph~R. Ghilardi,
  Jonathan~P. Lee, Patrick~W. Mantyh, and John~E. Maggio.
\newblock Alzheimer's disease amyloid propagation by a template-dependent
  dock-lock mechanism.
\newblock {\em Biochemistry}, 39(21):6288--6295, May 2000.

\bibitem{Meisl2014}
Georg Meisl, Xiaoting Yang, Erik Hellstrand, Birgitta Frohm, Julius~B.
  Kirkegaard, Samuel I.~A. Cohen, Christopher~M. Dobson, Sara Linse, and Tuomas
  P.~J. Knowles.
\newblock Differences in nucleation behavior underlie the contrasting
  aggregation kinetics of the a$\beta$40 and a$\beta$42 peptides.
\newblock {\em Proceedings of the National Academy of Sciences},
  111:9384--9389, 2014.

\bibitem{Cohen2013}
Samuel I.~A. Cohen, Sara Linse, Leila~M. Luheshi, Erik Hellstrand, Duncan~A.
  White, Luke Rajah, Daniel~E. Otzen, Michele Vendruscolo, Christopher~M.
  Dobson, and Tuomas P.~J. Knowles.
\newblock Proliferation of amyloid-beta42 aggregates occurs through a secondary
  nucleation mechanism.
\newblock {\em Proceedings of the National Academy of Sciences},
  110:9758--9763, 2013.

\bibitem{Debye1923}
Peter~Debye and Erich~H\"uckel.
\newblock Zur theorie der elektrolyte. i. gefrierpunktserniedrigung und
  verwandte erscheinungen.
\newblock {\em Phys.~Z}, 24:185, 1923.

\bibitem{Bronsted1924}
Johannes~N. Bronsted and C.~E. Teeter.
\newblock On kinetic salt effect.
\newblock {\em The Journal of Physical Chemistry}, 28(6):579--587, 1923.

\bibitem{Meisl2016a}
Georg Meisl, Xiaoting Yang, Birgitta Frohm, Tuomas P.~J. Knowles, and Sara
  Linse.
\newblock Quantitative analysis of intrinsic and extrinsic factors in the
  aggregation mechanism of alzheimer-associated a$\beta$-peptide.
\newblock {\em Scientific Reports}, 6:18728, 2016.

\bibitem{Szczepankiewicz2015}
Olga Szczepankiewicz, Björn Linse, Georg Meisl, Eva Thulin, Birgitta Frohm,
  Carlo~Sala Frigerio, Michael~T. Colvin, Angela~C. Jacavone, Robert~G.
  Griffin, Tuomas Knowles, Dominic~M. Walsh, and Sara Linse.
\newblock N-terminal extensions retard a$\beta$42 fibril formation but allow
  cross-seeding and coaggregation with a$\beta$42.
\newblock {\em Journal of the American Chemical Society}, 137(46):14673--14685,
  2015.

\bibitem{Bolognesi2014}
Benedetta Bolognesi, Samuel I.~A. Cohen, Pablo~Aran Terol, Elin~K. Esbjörner,
  Sofia Giorgetti, Maria~F. Mossuto, Antonino Natalello, Ann-Christin Brorsson,
  Tuomas P.~J. Knowles, Christopher~M. Dobson, and Leila~M. Luheshi.
\newblock Single point mutations induce a switch in the molecular mechanism of
  the aggregation of the alzheimer’s disease associated aβ42 peptide.
\newblock {\em ACS Chemical Biology}, 9(2):378--382, 2014.

\bibitem{Walsh2002}
Dominic~M. Walsh, Igor Klyubin, Julia~V. Fadeeva, William~K. Cullen, Roger Anwyl,
  Michael~S. Wolfe, Michael~J. Rowan, and Dennis~J. Selkoe.
\newblock Naturally secreted oligomers of amyloid beta protein potently inhibit
  hippocampal long-term potentiation in vivo.
\newblock {\em Nature}, 416(6880):535--539, Apr 2002.

\bibitem{Jan2011}
Asad Jan, Oskar Adolfsson, Igor Allaman, Anna-Lucia Buccarello, Pierre~J.
  Magistretti, Andrea Pfeifer, Andreas Muhs, and Hilal~A. Lashuel.
\newblock A$\beta$42 neurotoxicity is mediated by ongoing nucleated
  polymerization process rather than by discrete a$\beta$42 species.
\newblock {\em Journal of Biological Chemistry}, 286(10):8585--8596, 2011.

\bibitem{Walsh2009}
Dominic~M. Walsh, Eva~Thulin, Aedin~M. Minogue, Niklas~Gustavsson, Eric Pang, David~B. Teplow, and
  Sara Linse.
\newblock A facile method for expression and purification of the alzheimer's
  disease-associated amyloid beta-peptide.
\newblock {\em FEBS J}, 276(5):1266--1281, Mar 2009.

\bibitem{Wales1997}
David~J. Wales and Jonathan P.~K. Doye.
\newblock Global optimization by basin-hopping and the lowest energy structures
  of lennard-jones clusters containing up to 110 atoms.
\newblock {\em The Journal of Physical Chemistry A}, 101(28):5111--5116, 1997.

\end{thebibliography}
\end{document}

\maketitle

\renewcommand{\thefigure}{S\arabic{figure}}
\renewcommand{\theequation}{S\arabic{equation}}
\renewcommand{\thepage}{S\arabic{page}}   %adds S in front of figure, equ and page number

\section{Additional fits and datasets}
\subsection{Full datasets from Fig.~4}%\ref{fig:2mM_fit}
\begin{figure}
\includegraphics[width=\columnwidth]{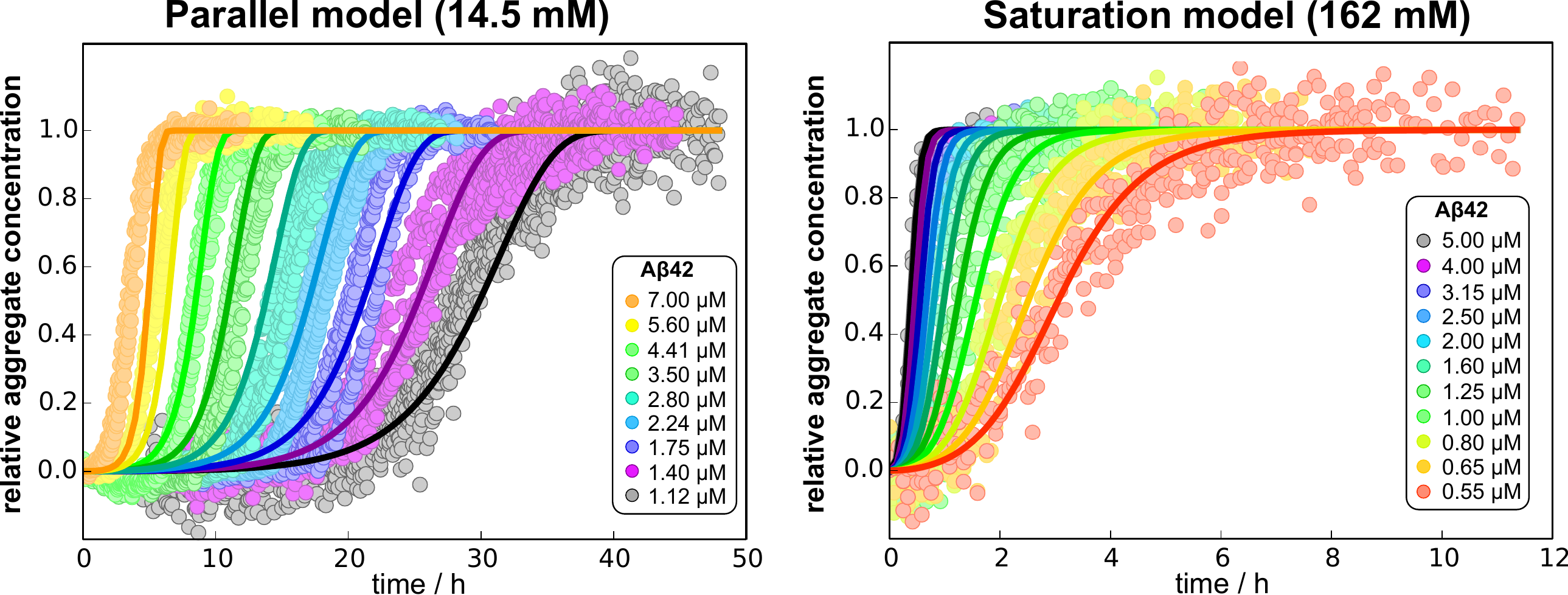}
\caption{\textbf{Fits from Fig.~4%\ref{fig:2mM_fit}
.} In Fig.~4%\ref{fig:2mM_fit}
 only part of the dataset was shown, here we display all monomer concentrations for the fits of the parallel model to the data at 2.5 mM salt and the fits of the saturation model to the data at 150 mM salt.}
\label{fig:2mM_150mM_fullfits}
\end{figure}

For clarity we showed only part of the dataset in the fits in Fig.~4%\ref{fig:2mM_fit}
of the main text. The full datasets are shown here in Fig.~\ref{fig:2mM_150mM_fullfits}. Note that the fits shown in the main text were also performed on the entire dataset, i.e. they are the same as the ones shown here, some concentrations were simply omitted for clarity.

\subsection{Fitting reaction orders}
\begin{figure}
\includegraphics[width=\columnwidth]{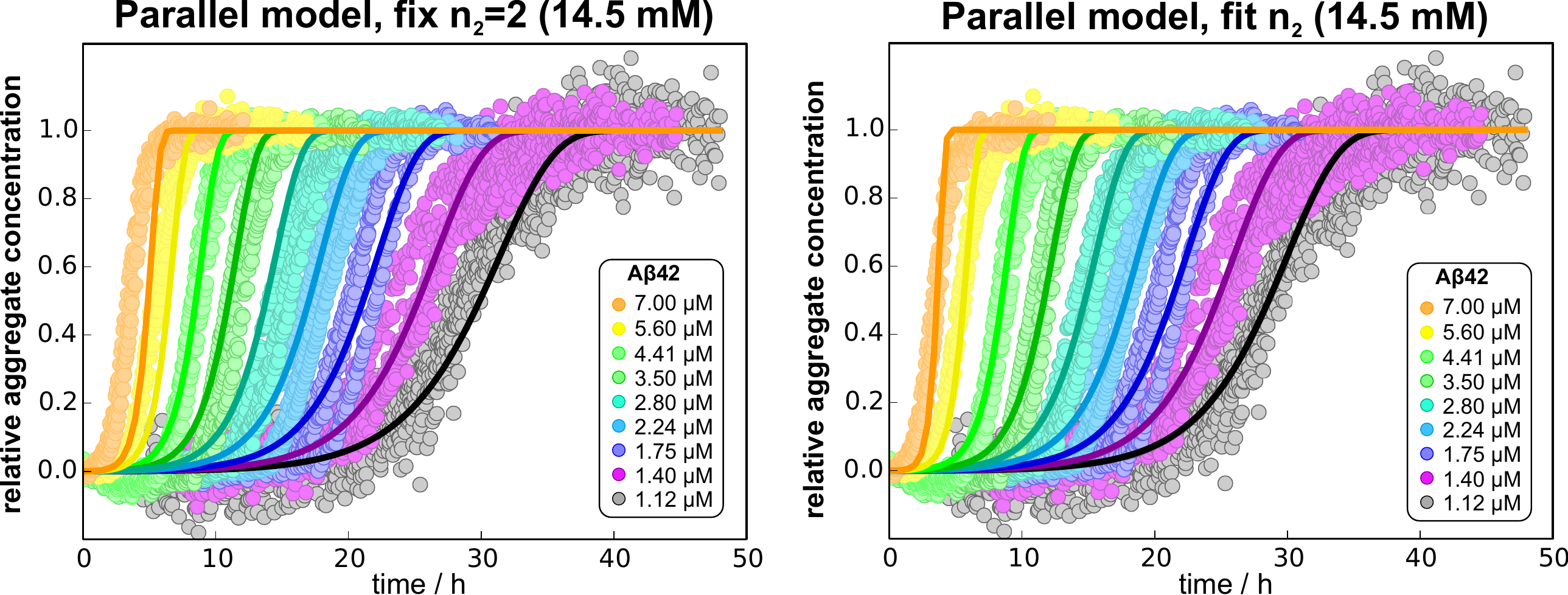}
\caption{\textbf{Fit of the full kinetic curves at 2.5 mM salt with $n_2$ as a free parameter.} On the left the secondary nucleation reaction order,$n_2$, is fixed to 2, which is the case used in the main text, on the right $n_2$ is left to vary during the fitting. $n_2$ converged to a value of 4, when left to vary, yielding slightly improved fits compared to the case of $n_2=2$.}
\label{fig:2mM_n2free_fit}
\end{figure}

In the main text we noted a slight deviation of the parallel model in both half times and fits at high monomer concentrations, Fig.~4%\ref{fig:2mM_fit}
. This may in fact be due to a small increase in the reaction order of secondary nucleation, e.g. due to an increase in secondary nucleus size. A freely varying $n_2$ achieved fits to the experimental data within error, as shown in Fig.~\ref{fig:2mM_n2free_fit}, with a value of $n2\approx4$.  However, the improvement in the fit is small and all other salt concentrations fit well to a value of $n2=2$. In addition a constant value of $n_2$ across the salt concentration allows for better comparison of the fitted rates. In the main text we therefore chose to keep $n_2$ fixed to a value of 2.

\newpage

\subsection{Detailed variation of rates with salt for each model.}
\begin{figure}[h]
\includegraphics[width=0.8\columnwidth]{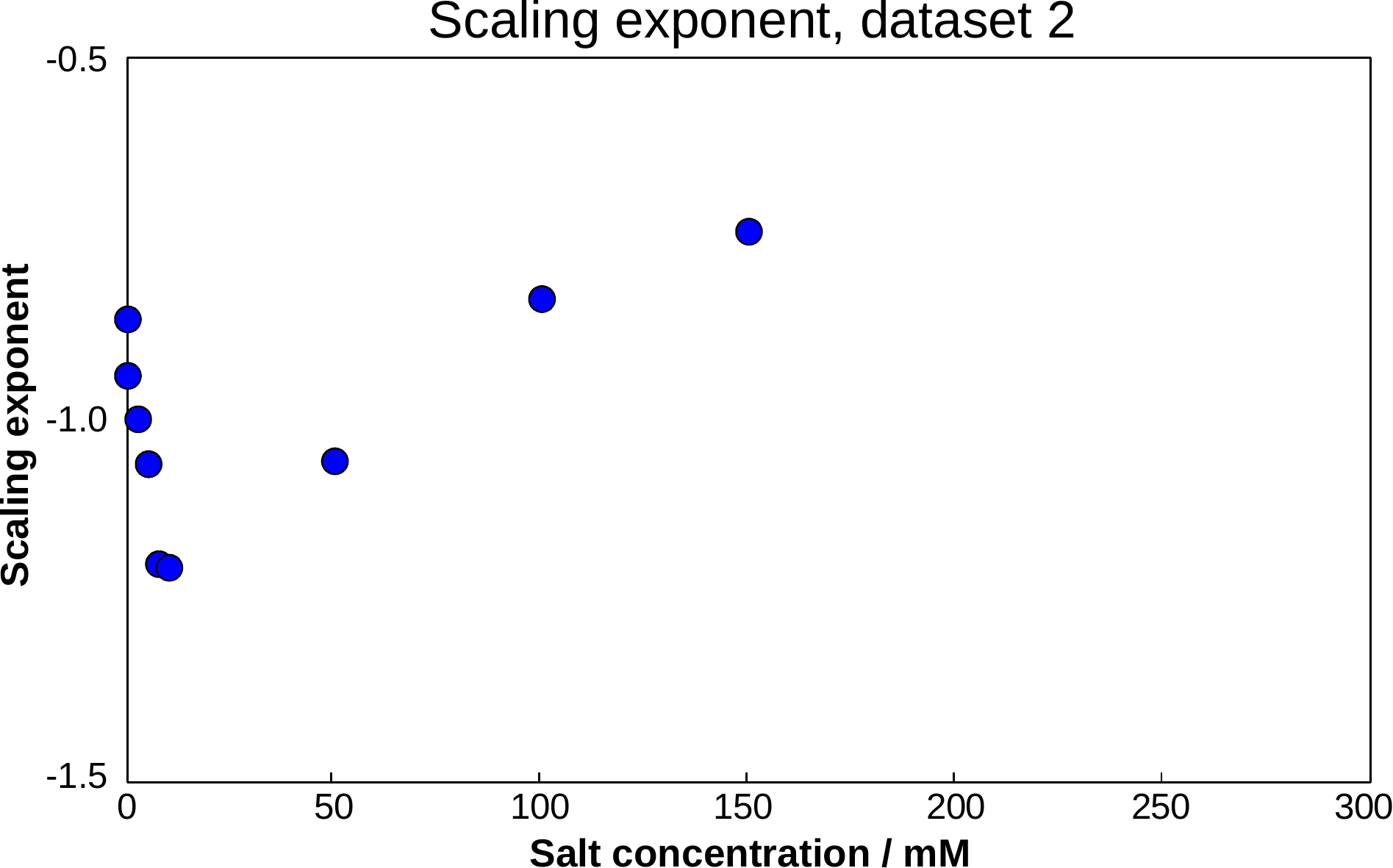}
\centering
\caption{\textbf{Variation of the average scaling exponent.} This is equivalent to the plot in Fig.~2%\ref{fig:halftimes_salt}
c in the main text, for the repeat dataset, and shows very similar results.}
\label{fig:scaling_exp}
\end{figure}

\begin{figure}
\includegraphics[width=\columnwidth]{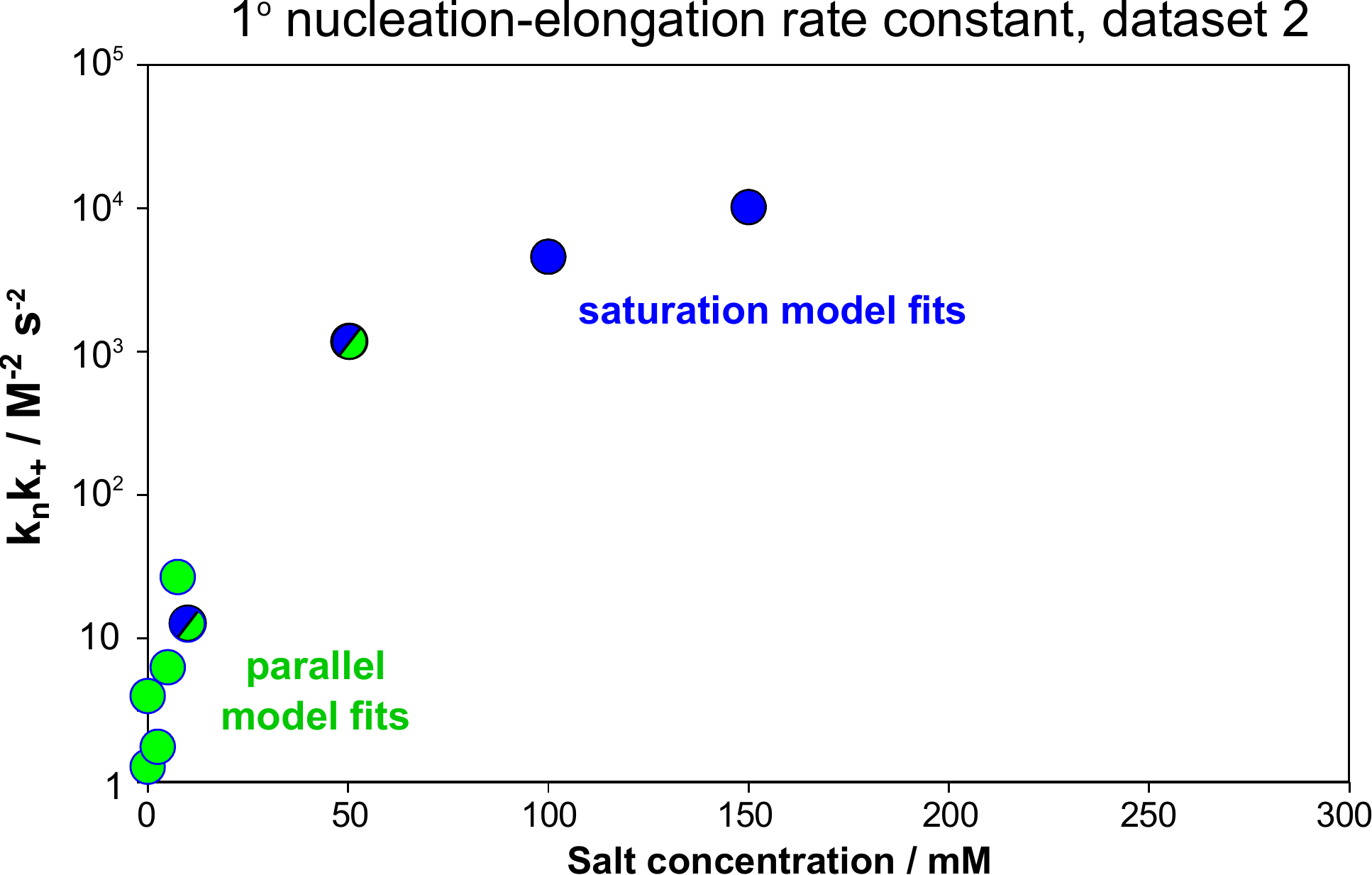}
\centering
\caption{\textbf{Variation of the combined primary nucleation-elongation rate constant.} This is the equivalent of Fig.~6%\ref{fig:ratio_rates}
b in the main text, for the repeat dataset, and shows very similar results.}
\label{fig:kpkn_set2}
\end{figure}

\begin{figure}
\includegraphics[width=\columnwidth]{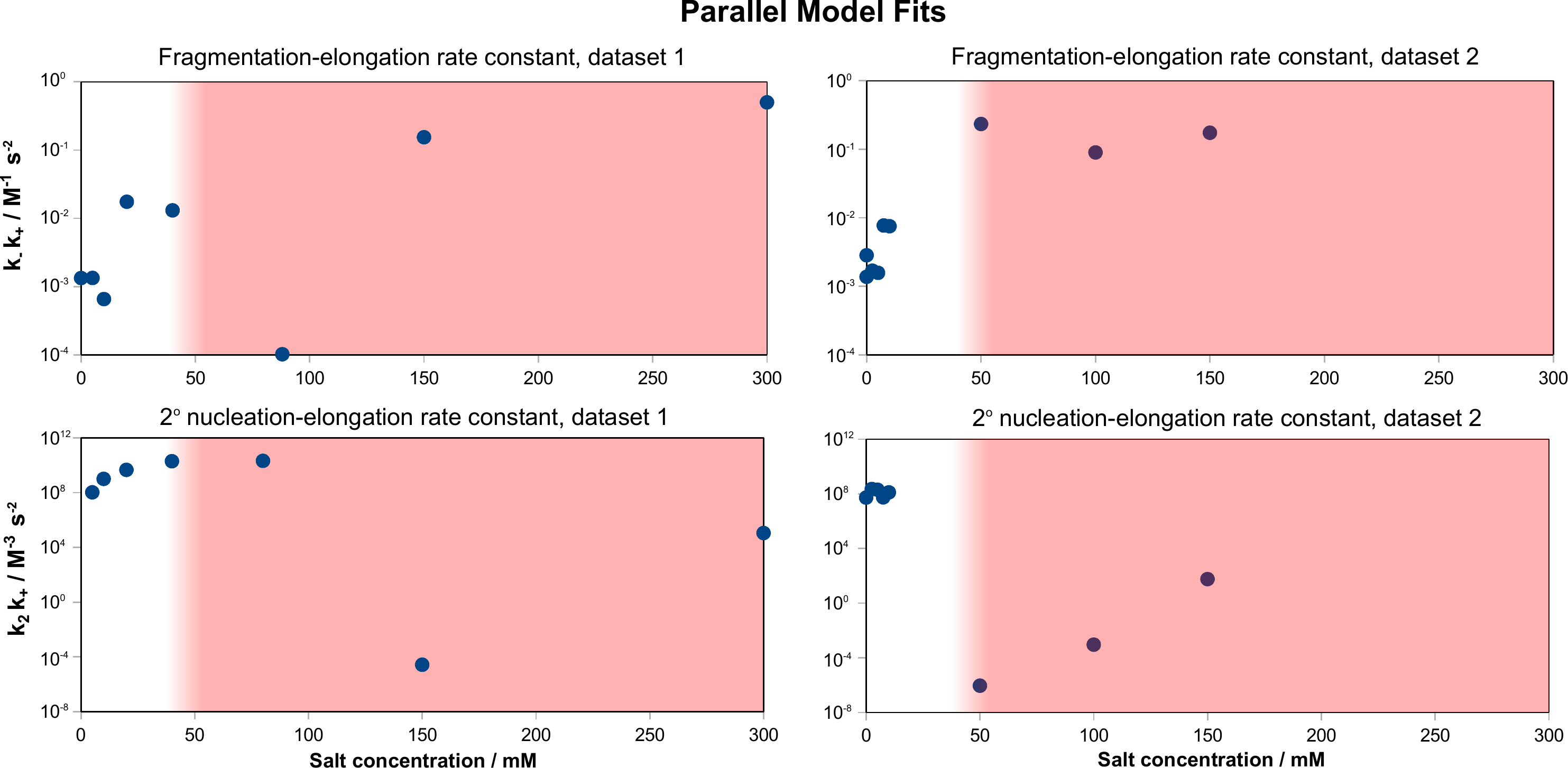}
\caption{\textbf{Variation of the microscopic rates with salt concentration for the parallel model.} The region in which the fragmentation model is no longer valid (above approximately 50 mM salt) is marked in red. The top panels show the combined fragmentation-elongation rate constant. The bottom panels show the combined secondary nucleation - elongation rate. Note the major discontinuity in $k_+k_2$ (15 orders of magnitude), which marks the breakdown of the parallel model, i.e. when saturated secondary nucleation is suddenly interpreted as fragmentation as explained in detail in the text. In the region of validity, $k_+k_2$ increases by several orders of magnitude, whereas $k_+k_-$ only increases by one order of magnitude. As the elongation rate constant was independently determined to increase by one order of magnitude, the increase in $k_+k_-$ can hence be  attributed purely to the increase in elongation. Therefore the fragmentation rate depends only weakly on the salt concentration, whereas the secondary nucleation rate constant (also confirmed by the saturation model) increases significantly with increasing salt concentration.}
\label{fig:rates_parallel}
\end{figure}

\begin{figure}
\includegraphics[width=\columnwidth]{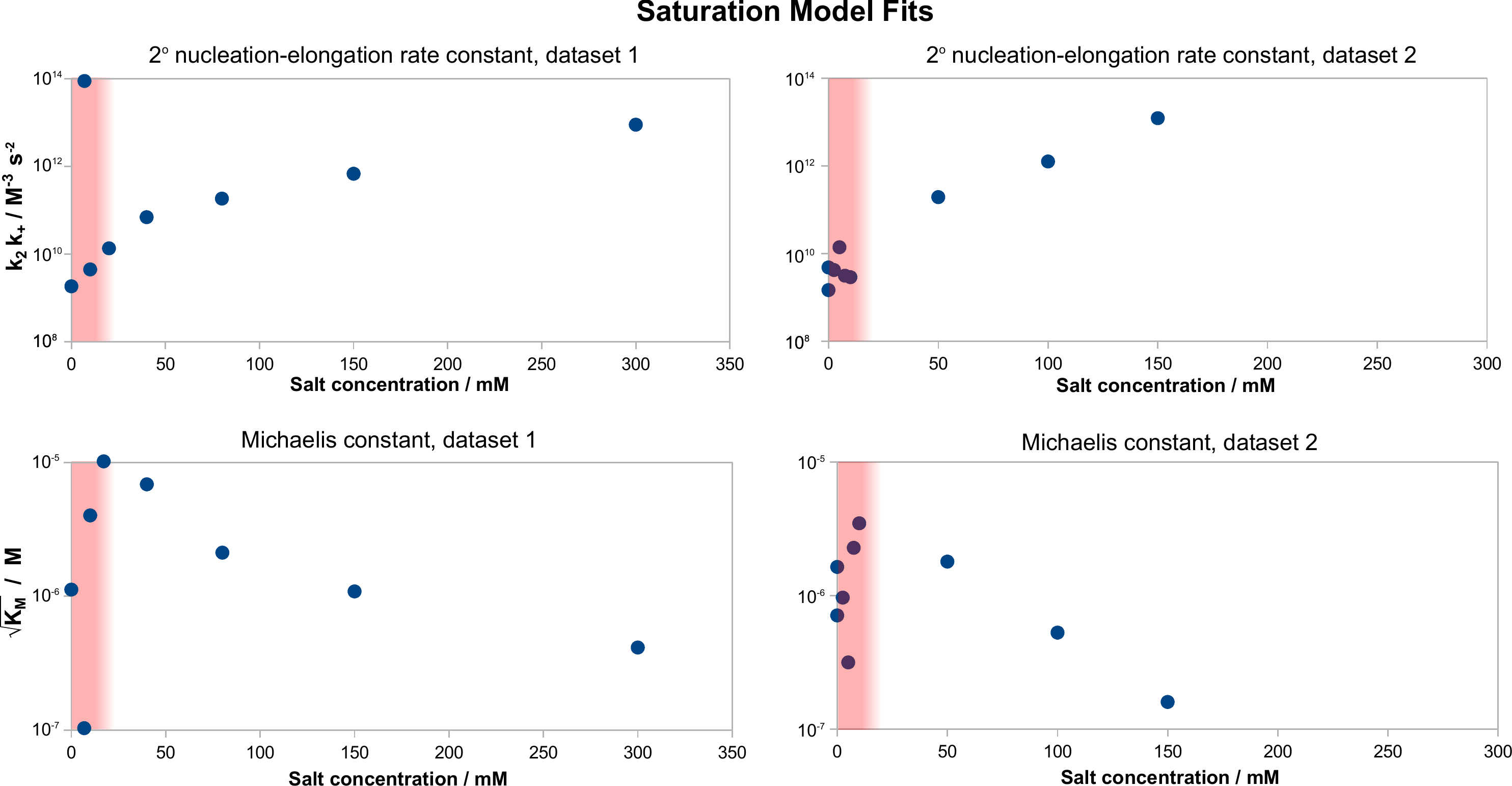}
\caption{\textbf{Variation of the microscopic rates with salt concentration for the saturation model.} The region in which the saturation model is no longer valid (below approximately 10 mM salt) is marked in red. The top panels show the combined secondary nucleation-elongation rate constant. The bottom panels show the square root of the Michaelis constant, i.e. the concentration at which saturation effects are most significant. In its region of validity, the secondary nucelation rate constant increase by about 2 orders of magnitude, the elongation rate constant, which was determined independently, shows only a weak increase in this region, hence secondary nucleation is clearly favoured by an increase in salt concentration. The Michaelis constant displays erratic behaviour in the region where the saturation model is not valid, as these values were obtained from missfits. In its region of validity there is a clear trend going from an unsaturated system, with saturation concentrations above the highest sampled monomer concentrations, to a saturated system, with saturation concentrations below the lowest sampled monomer concentrations.}
\label{fig:rates_saturation}
\end{figure}

The entire experimental setup was repeated and Fig.~\ref{fig:scaling_exp} shows the scaling exponent obtained from the second set of experiments and is in agreement with the findings of the first set of experiments.
In Figs.~\ref{fig:kpkn_set2}, \ref{fig:rates_parallel} and \ref{fig:rates_saturation}, the rates from fits of both models at all salt concentrations are shown. The entire setup was performed twice, labelled set 1 and set 2 in the figures, the results from the two datasets agree well.  By considering the saturation of secondary nucleation at high salt we can now explain the discontinuity observed in the parameters for the parallel model: At high salt concentrations the secondary nucleation process is saturated at all monomer concentrations. In a fully saturated system the term for secondary nucleation has no monomer dependence and becomes $k_2\frac{m(t)^{n_2}}{1+m(t)^{n_2}/K_M}M(t)\rightarrow k_2K_MM(t)$ (details see section "`Limits of models"' ) which is equivalent to a system purely determined by fragmentation, with fragmentation rate constant $k_-=k_2K_M$. Hence the parallel model yields good, purely fragmentation dominated fits at high salt concentrations, but the fragmentation rate constant obtained from the fit is in fact the saturated secondary nucleation rate constant, meaning that the discontinuities in $k_-$ and $k_2$  in the parallel model results from the fact that we are fitting to actual fragmentation at low salt concentrations and misinterpreting saturated secondary nucleation as fragmentation at high salt concentrations.

\newpage

\section{Theoretical Background and Detailed Solutions}
\paragraph{Model derivation}
We make two basic assumptions, first that the increase in fibril mass is proportional to the number of fibrils (this is motivated by considering addition to the ends of fibrils only; the area and structure of the ends is not expected to change with fibril length) and that the processes creating new fibrils are proportional to the total fibril mass (what we will call secondary processes; this includes fibril surface catalysed nucleation and fragmentation of fibrils). 
Using simple chemical kinetics we can then derive differential equations, the moment equations, describing the time evolution of fibril number concentration, $P(t)$, and fibril mass concentration, $M(t)$. 
\begin{eqnarray}
\frac{dP}{dt}&=&k_n m(t)^{n_c}+k_-(M(t)-(2n_c-1)P[t]) +k_2 \frac{m(t)^{n_2}}{1+m(t)^{n_2}/K_M}M(t) \label{eq:dp_const_SI}\\
\frac{dM}{dt}&=&2 m(t)k_+P(t)
\label{eq:dP_dM_SI}
\end{eqnarray}
The origin of the terms in equation (\ref{eq:dp_const_SI}) from left to right is primary nucleation, fragmentation and secondary nucleation. In previous work \cite{Meisl2014} it was shown that secondary nucleation may saturate at high monomer concentrations, i.e. become monomer independent. This is reflected in the term $\frac{m(t)^{n_2}}{1+m(t)^{n_2}/K_M}$ which originates from a treatment of secondary nucleation as a Michaelis-Menten like process resulting in saturation kinetics: At low monomer concentrations secondary nucleation  depends on monomer to the power of $n_2$ ($\frac{m(t)^{n_2}}{1+m(t)^{n_2}/K_M}\approx m(t)^{n_2}$), whereas at high monomer concentrations it is independent of monomer concentration ($\frac{m(t)^{n_2}}{1+m(t)^{n_2}/K_M}\approx K_M$). The Michaelis constant $K_M$ determines at which concentration this change of behaviour occurs. 

The solution to these equations in the limit $K_M\rightarrow \infty$ (parallel model) is outlined below and in the limit $k_-\rightarrow 0$ (saturation model) it is found in "Differences in nucleation behavior underlie the contrasting aggregation kinetics of the A$\beta$40 and A$\beta$42 peptides" \cite{Meisl2014}. 

For the parallel model the system of differential equations to solve becomes:
\begin{eqnarray}
\frac{dP}{dt}&=&k_n m(t)^{n_c}+k_-M(t)+k_2m(t)^{n_2}M(t) \label{eq:dp_const_comp_SI}\\
\frac{dM}{dt}&=&2 m(t)k_+P(t)
\label{eq:dP_dM_comp_SI}
\end{eqnarray}
where we ignored the term $-(2n_c-1)P[t]$ for clarity. It accounts for the fibrils lost when a piece smaller than the critical nucleus breaks off and dissolves, which is negligible for the small nucleus size at hand.
We find the early time solution to these equations by setting $m(t)=m(0)$; this linearises the equations and makes them easily solvable. Then, by integrating equation \ref{eq:dp_const_comp_SI} we obtain a fixed point operator and substitute the early time solution as an initial guess into this fixed point operator to yield the first order solution for $M(t)$. This follows the strategy outlined in detail in Cohen \textit{et al.}(2011) \cite{Cohen2011a,Cohen2011b}.
The approximate solution obtained in this way is given by:

\begin{eqnarray}
M(t)&=&m_{\mathrm{tot}}+\mathrm{Exp}\left[-\frac{k_+(4c\kappa\mathrm{Cosh}(\kappa t)+4P_0\kappa^2\mathrm{Sinh}(\kappa t))}{2\kappa^3}\right] \nonumber \\ 
&&\left((M_0-m_{\mathrm{tot}})e^{\frac{2k_+c}{\kappa^2}}\right)
\label{eq:M_parallel_S}
\end{eqnarray}
where
\begin{eqnarray}
a&=&k_2m_0^{n_2}+k_- \nonumber \\
c&=&k_nm_0^{n_c}+aM_0 \nonumber \\
\kappa&=&\sqrt{2k_+m_0(k_2m_0^{n_2}+k_-)}
\end{eqnarray}
where $m_{\mathrm{tot}}$ is the total protein concentration, $M_0$, $P_0$ and $m_0$ are the initial mass concentration of fibrils, number concentration of fibrils and monomer concentration respectively, $k_+$, $k_-$, $k_n$ and $k_2$ are the rate constants of elongation, fragmentation, primary nucleation and secondary nucleation.

Note that for unseeded experiments the two limiting cases only involve 3 free parameters each ($k_+k_2$, $k_+k_n$ and $k_+k_-$ for the parallel model and $k_+k_2$, $k_+k_n$ and $K_M$ for the saturation model) which are sufficient to produce good global fits to all monomer concentrations at any given salt concentration.

For the parallel model the scaling exponents is given by

\begin{equation}
\gamma_{\mathrm{comp}}= \frac{d\log(t_{1/2})}{d\log(m(0))}\approx       -\frac{1}{2}\left(\frac{n_2}{1+K/m(0)^{n_2}}+1\right)
\label{eq:scaling_frag_secnuc_SI}
\end{equation}
where $K=k_-/k_2$. This interpolates between  $\gamma = -1/2$ and $\gamma = -(n_2+1)/2$ for the limits of low and high monomer respectively, giving the negative curvature in the double logarithmic plots of the half time, as predicted from the qualitative argument used to derive the general constraints above.

For the saturation model the scaling exponent is given by
\begin{equation}
\gamma_{\mathrm{sat}}\approx       -\frac{1}{2}\left(\frac{n_2}{1+m(0)^{n_2}/K_M}+1\right)
\label{eq:scaling_secnuclim_SI}
\end{equation}
where $K_M$ is the Michaelis constant. This interpolates between  $\gamma = -(n_2+1)/2$ and $\gamma = -1/2$  for the limits of low and high monomer respectively, i.e. the reverse of the other model.

\paragraph{Limits of models}
A fully saturated system in the saturation model ($m(t)^{n_2}\gg K_M$) is mathematically equivalent to a purely fragmentation dominated system in the parallel model ($k_-\gg k_2m(t)^{n_2}$) as they both include a single, monomer-independent secondary process. This explains why reasonable fits of the parallel model were achieved at high salt concentrations, albeit with a discontinuity in the fitted rates compared to lower salt, because saturated secondary nucleation was missinterpreted as fragmentation by the fitting, leading to unrealistic behaviours of the associated rates. 

Equally, a completely unsaturated system in the saturation model ($m(t)^{n_2}\ll K_M$) is mathematically equivalent to a purely secondary nucleation dominated system in the parallel model ($k_-\ll k_2m(t)^{n_2}$) as they both include one single-step, monomer-dependent secondary process.
However, because the analytical solutions given above are approximate, they do in fact not converge to exactly the same solutions in these limits. The solution to the parallel model is obtained through one fixed point iteration of the early time linearised solutions, whereas the saturation model is obtained from one fixed point iteration of a slightly improved initial guess \cite{Cohen2011a,Cohen2011b, Meisl2014}, meaning that the saturation model analytical solutions are more accurate, although this effect is minor under most conditions. However this may be one of the reasons the fits in the region where both models are valid do not agree perfectly. 

\newpage

\section{Seeded experiments and analysis}
\subsection{Low concentration seeding}
\begin{figure}[h]
\includegraphics[width=\columnwidth]{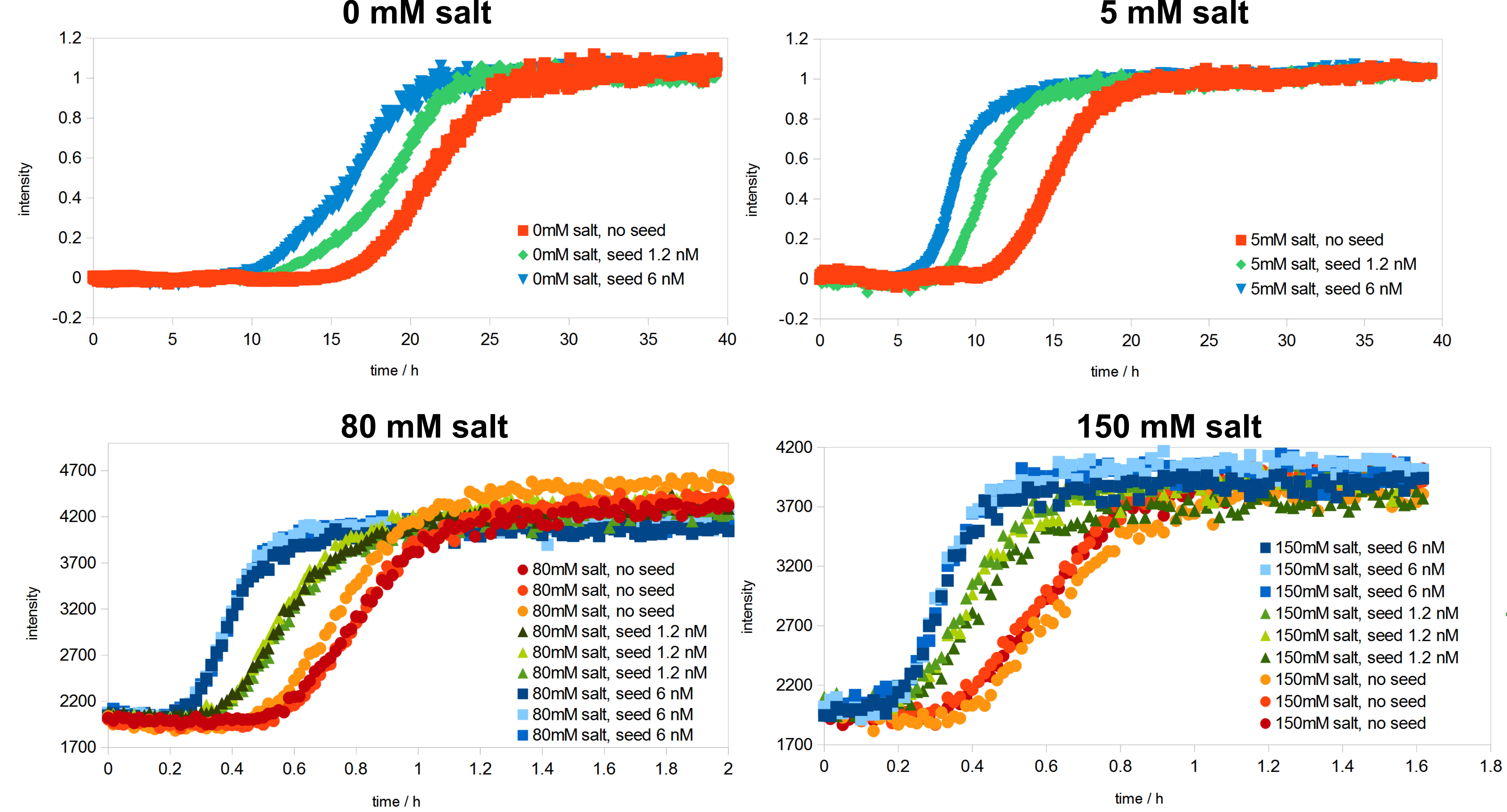}
\caption{\textbf{Effect of low seeding at different salt.} The aggregation of 3 $\mu$M A$\beta$42 at a number of salt concentrations between 0 mM and 150 mM NaCl. The curves at low salt are normalised and averaged over triplicate repeats for reasons of clarity as the variations at these concentrations are large. For the higher salt concentrations the individual curves are shown. In all cases 1.2 nM seeds, corresponding to 0.04\% of the monomer concentration present, already lead to a significant shortening of the half time, confirming secondary nucleation is dominant.}
\label{fig:low_seed}
\end{figure}

In order to confirm that secondary processes do indeed dominate at all salt concentrations, we performed seeded experiments at very low seed concentrations (nM). If the system is dominated by primary nucleation this addition of a very small concentration of seed material has little effect and the half times are expected to remain unchanged. However, if the system is dominated by secondary nucleation these seeds catalyse the formation of more seeds in the positive feedback mechanism of secondary nucleation. This leads to a significant effect on the half times as is apparent at all tested salt concentration, see Fig.~\ref{fig:low_seed}.

\subsection{Determination of elongation rate constant}
\begin{figure}[h]
\includegraphics[width=\columnwidth]{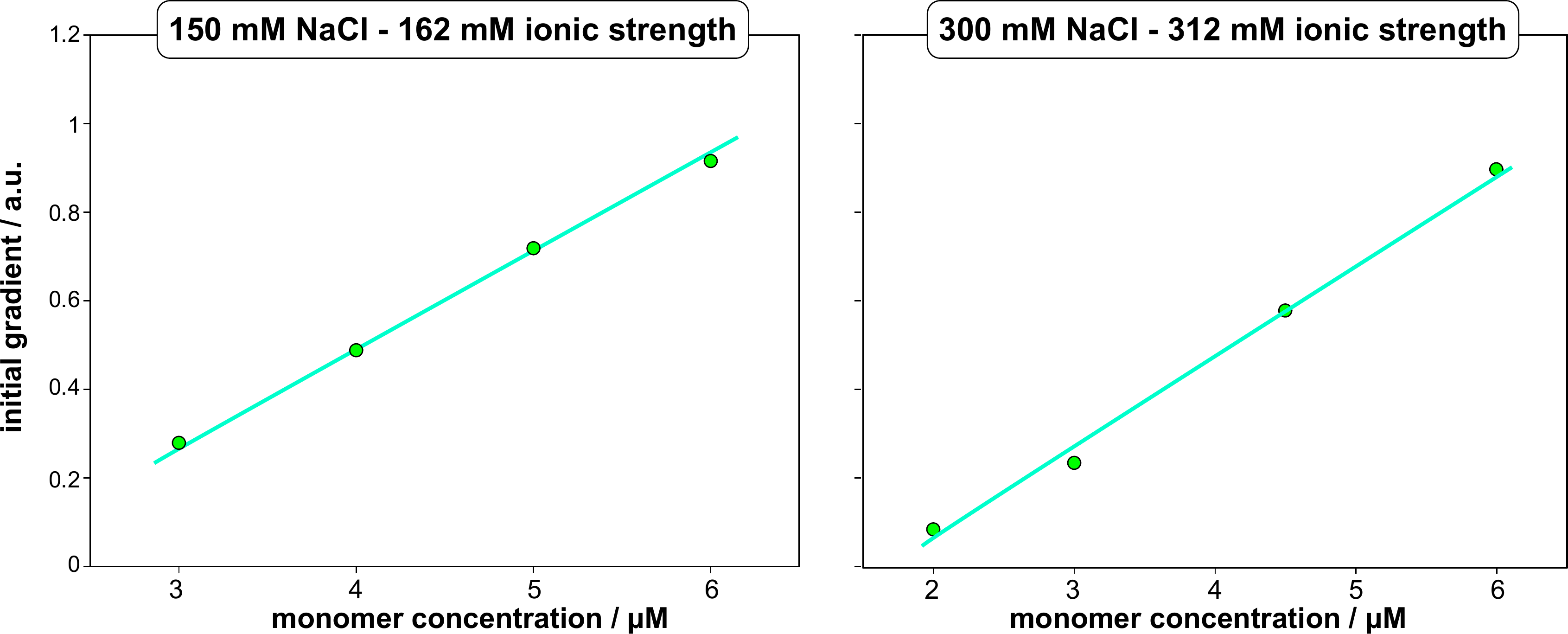}
\caption{\textbf{Initial gradients from strongly seeded experiments.} The initial gradients from strongly seeded experiments (2 $\mu$M seeds), show a linear dependence on the monomer concentration, even at high ionic strengths, suggesting that the elongation rate is not yet beginning to saturate. The factor of proportionality relating the initial gradient to the monomer concentration can be used to determine the elongation rate constant.}
\label{fig:init_grad}
\end{figure}

\begin{figure}[h]
\includegraphics[width=\columnwidth]{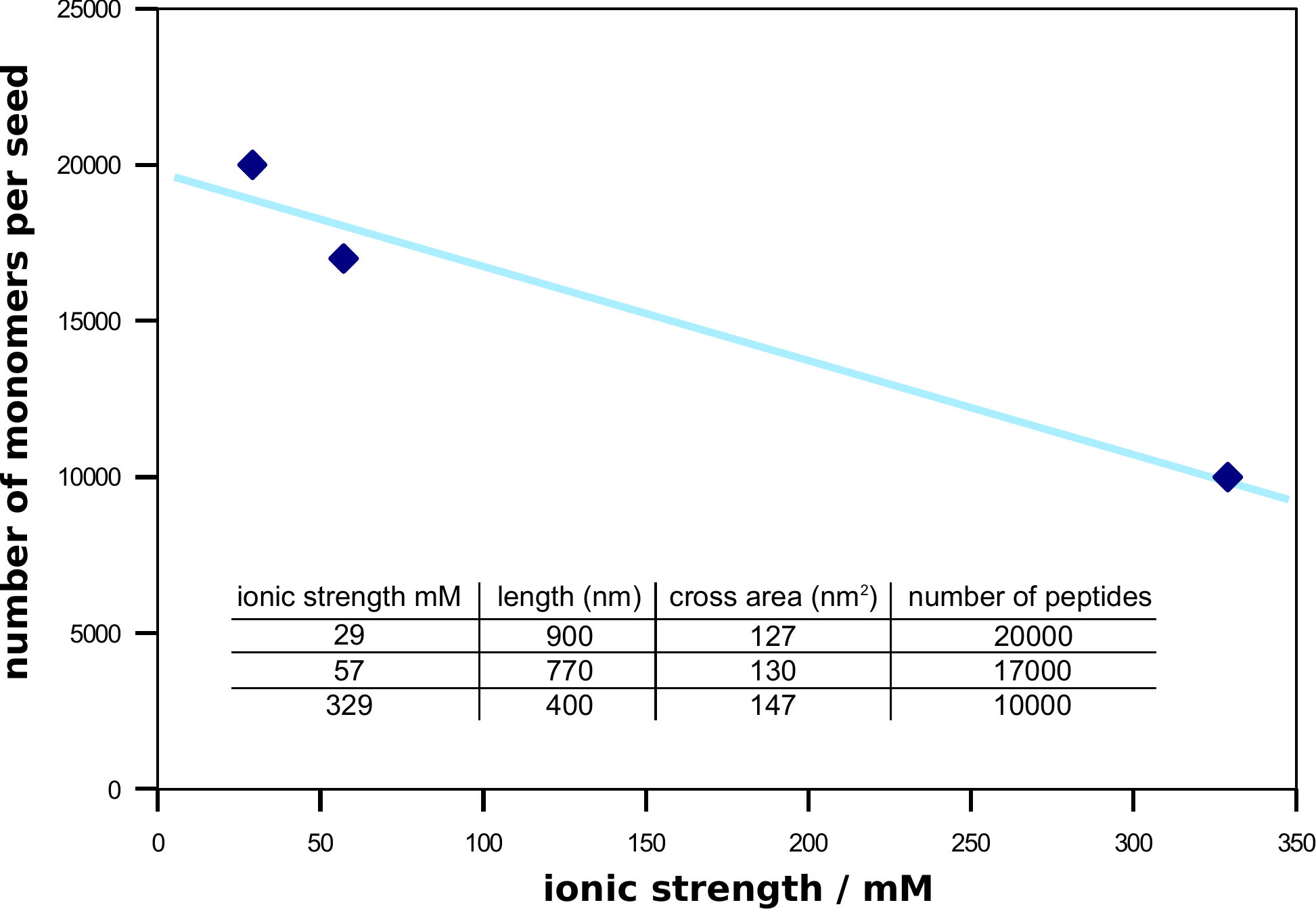}
\caption{\textbf{Determination of fibril dimensions from TEM.} A slight decrease in fibril length upon increase in ionic strength can be observed. This trend is in agreement with our finding that the elongation rate increases less than the nucleation rates, leading to a shorter equilibrium fibril length. }
\label{fig:TEM_lengths}
\end{figure}

The estimation of the elongation rate also follows the methods described in \cite{Meisl2014}: Strongly seeded experiments, at a seed concentration of 2 $\mu$M and a monomer concentrations ranging from 1 to 6 $\mu$M were performed at salt concentrations of 1, 10, 31, 80 ,150, and 300 mM (ionic strengths are 12 mM higher due to the presence of buffer). The initial gradient, $dP/dt|_{t=0}=2k_+P_0m(t)$, at the different monomer concentrations was used to extract the constant of proportionality, $2k_+P_0$, relating the initial rate to the monomer concentration.
This moreover serves as a check for saturation of the elongation rate: If the elongation rate does saturate, then the increase of the initial gradient with monomer concentration is not linear, but plateaus above a certain monomer concentration. No evidence of such a plateau was observed, even at the highest salt concentrations, as evident in Fig.~\ref{fig:init_grad}.
In order to determine the absolute value of the elongation rate constant, $k_+$, from this, $P_0$ needs to be estimated. The average number of monomers per seed fibril, $L_0$, relates the initial fibril mass concentration, $M_0$, to the initial fibril number concentration, $P_0$, by $P_0=M_0/L_0$. The fibril mass concentration is known and $L_0$ can be determined from a measurement of fibril dimensions by TEM. Fig.~\ref{fig:TEM_lengths} shows a table of the average dimensions of fibrils at the different salt concentrations. By assuming a density of protein of 1.3 g/ml these dimensions can be used in order to calculate the number of monomers per seed fibril, which is plotted in Fig.~\ref{fig:TEM_lengths}. The absolute values are likely to be very approximate, due to various biases in the extraction of fibril lengths: larger fibrils are harder to detect, as the likelihood of being able to detect both ends is smaller. Moreover fibrils can form tangles, which will make the ends less accessible and thereby increase the number of fibrils per growth competent end, $L_0$. However, if we assume this effects similarly affect all salt concentrations, the observed trend will be more reliable than the absolute values. This equally translates to the trend of the elongation rate constant with increasing salt: although its absolute value is inaccurate due to the errors associated in determining the seed length, its increase by an order of magnitude is significant. The error bars shown in Fig.~6%\ref{fig:ratio_rates}
 are hence determined by the inaccuracies in measuring the initial gradient and in extrapolating the seed length to salt concentrations not measured explicitly (Fig.~\ref{fig:TEM_lengths}).

\newpage

\section{Further experimental and theoretical techniques}
\subsection{Error estimation}
The error estimation was based on the method described in \cite{Meisl2014}: All experiments were performed as triplicate repeats, therefore in order to obtain an estimate of the error in the rate constants due to experimental variations, we considered subsets of the data, including one repeat only at each monomer concentration. These subsets were fitted normally and the resulting values of the rate constants were averaged and their standard deviation used to obtain an estimate of the error.

\subsection{Fitting algorithm}
The fitting was performed using our protein aggregation fitting software, AmyloFit\cite{Meisl2016}, which can be found at www.amylofit.ch.cam.ac.uk. It employs a basin hopping algorithm \cite{Wales1997}, which consists of a local minimisation paired with a Monte-Carlo step to randomise initial guesses. This algorithm thereby effectively flattens the potential energy landscape and allows the fit to converge to the global minimum even if the initial guess was not in the correct basin of attraction. Such a property is important in the global fitting of large datasets to complex equations as these situations tend to result in rough energy landscapes and convergence to local rather than the global minimum can be an issue.